\documentclass[aps,pra,reprint,superscriptaddress,longbibliography,floatfix,raggedbottom]{revtex4-2}
\usepackage[T1]{fontenc}
\usepackage{lmodern}
\usepackage{amsmath,amssymb,graphicx}
\usepackage{booktabs}
\usepackage{hyperref}
\hypersetup{
  hidelinks,
  breaklinks=true,
  pdftitle={Extensively degenerate superradiant phase in a giant-atom quantum Rabi ring},
  pdfauthor={Shu He and Li-Wei Duan}
}
\begin{document}
\title{Extensively degenerate superradiant phase in a giant-atom quantum Rabi ring}
\author{Shu He}
\email{heshu1987@foxmail.com}
\affiliation{Department of Physics and Electronic Engineering, Sichuan Normal University, Chengdu 610066, China}
\author{Li-Wei Duan}
\email{duanlw@gmail.com}
\affiliation{Department of Physics, Zhejiang Normal University, Jinhua 321004, China}
\begin{abstract}
We show that an $N$-site quantum Rabi ring with matched giant-atom couplings has $2^N$ degenerate superradiant configurations as the atomic-to-cavity frequency ratio $\eta=\Delta/\omega$ tends to infinity. Each atom couples to two cavities separated by two lattice spacings, and the photon hopping carries a phase $\pi/2$. Matching the ratio of the two coupling strengths to the dimensionless photon hopping reduces the mean-field energy to a sum of single-site Rabi energies, with a common critical coupling for all momentum-pair sectors. The degeneracy survives quadratic quantum fluctuations despite a dispersive excitation spectrum. At large but finite $\eta$, second-order processes involving the local cubic anharmonicity generate a same-sublattice Ising coupling of order $\omega/\eta$, selecting ferro- and antiferro-superradiant configurations. This quantum order-by-disorder effect has no classical counterpart. Four-site calculations retaining all spin states quantitatively support the predicted energy coefficients. The separation between the induced coupling and the excitation gap allows a broad temperature interval with macroscopic photon occupation and short-range correlations between displacement signs.
\end{abstract}
\maketitle\raggedbottom
\section{Introduction}

The quantum Rabi model undergoes a superradiant phase transition when the atomic-to-cavity frequency ratio $\eta=\Delta/\omega$ tends to infinity, a limit that plays the role of the thermodynamic limit~\cite{Ashhab2013,Hwang2015,Liu2017}. Its critical behavior has been observed with a single trapped ion~\cite{Cai2021}. When photons hop between cavities, as in quantum Rabi lattices and rings~\cite{Schiro2012,Flottat2016,Mei2022}, the transition also selects a spatial configuration of the photon displacements. The sign and phase of the hopping determine whether this configuration is uniform, staggered, chiral, or frustrated~\cite{Zhang2021,FallasPadilla2022,Zhao2022,Li2023,Zhao2023,Qin2024}. The resulting ordered phases have a finite degeneracy, for example twofold in a ferro-superradiant phase and sixfold in the frustrated phase of a Dicke trimer~\cite{Zhao2022}.

Giant atoms provide additional control over light--matter interactions. An atom coupled to the field at spatially separated coupling points (contacts) experiences interference that can modify relaxation, generate bound states, and mediate interactions~\cite{Kockum2014,Kockum2018,Kannan2020,Wang2021,Du2022,TerradasBrianso2022,Wang2024}. In a Rabi lattice the same interference reshapes the static photon response that couples the displacements seen by different atoms. This raises the question of whether the contacts can be arranged so that all choices of the local displacement signs have the same mean-field energy while the photon band remains dispersive. Such a degeneracy would arise from contact design, rather than from flat-band localization in light--matter lattices~\cite{Zhu2020,He2021} or from the competing structures and photon glasses of multimode cavities~\cite{Gopalakrishnan2009,Strack2011,Buchhold2013,Vaidya2018,Guo2019}.

We show that a ring in which each atom couples to two cavities separated by two lattice spacings, with a hopping phase $\pi/2$, realizes this degeneracy. When the ratio of the two coupling strengths is matched to the dimensionless hopping, the mean-field energy becomes a sum of single-site Rabi energies. Each of the $N$ contact fields then independently takes either sign, giving $2^N$ degenerate superradiant configurations. All of them share the same quadratic Hamiltonian and zero-point energy, although the common excitation spectrum is dispersive. In the large-$\eta$ expansion~\cite{Hwang2015,Liu2017,Ye2025}, the two local minima have opposite cubic terms, and virtual three-excitation processes generate an Ising interaction of order $\omega/\eta$ between displacement signs on the same sublattice, selecting ferro- and antiferro-superradiant configurations. This is an example of quantum order by disorder~\cite{Harris1992,Chubukov1992,Barnett2012,Yue2021}, with selection arising from anharmonic terms as in the kagome antiferromagnet~\cite{Chernyshev2014}. Related cavity systems exhibit vacuum-induced phase selection~\cite{Schuler2020} and competition between quantized cavity fields and order by disorder~\cite{Liang2026}. In the present model, contact matching both produces the degeneracy and makes the classical distribution of the contact fields factorize exactly, so every sign configuration has equal classical weight and the selection is purely quantum.

The induced interaction is much smaller than the excitation gap about each configuration, which remains of order $\omega$ as $\eta$ increases. Between these two scales lies a temperature interval in which correlations between displacement signs are short ranged while the photon occupation remains of order $\eta$ per site. The loss of sign correlations is visible in the momentum distribution of the photons and raises the local photon number toward its value for uncorrelated signs. Four-site calculations retaining all spin states support the predicted energy coefficients and show that tunneling between configurations is suppressed exponentially in $\eta$.

\section{Model and mean-field degeneracy}\label{sec:model}

\begin{figure*}[t]
\includegraphics[width=\textwidth]{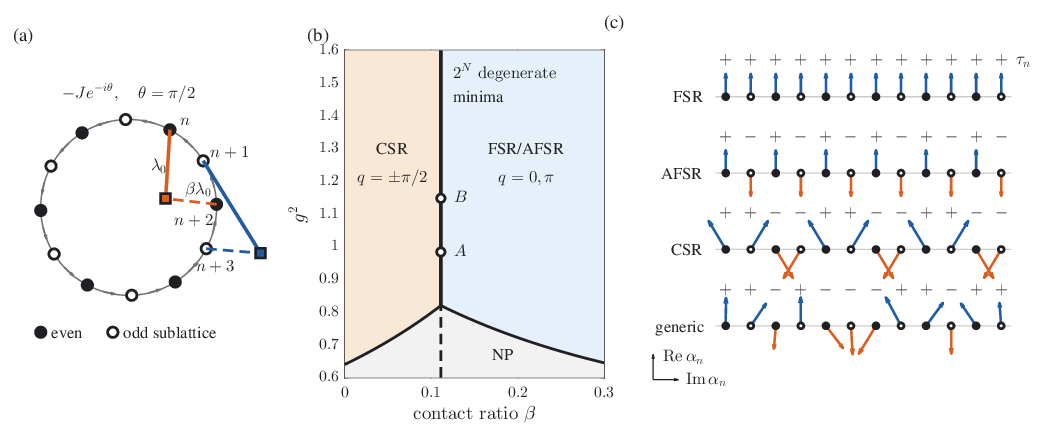}
\caption{Model, mean-field phase diagram, and degenerate configurations at $\theta=\pi/2$ and $j=0.6$, for $\eta\to\infty$. (a) Each atom (square) couples to cavities $n$ and $n+2$ with strengths $\lambda_0=\lambda/\sqrt{1+\beta^2}$ (solid) and $\beta\lambda_0$ (dashed); only two atoms are shown. Filled and open circles mark the even and odd sublattices. The photon hopping matrix element along each arrow, from $n+1$ to $n$, is $-Je^{-i\theta}$ [Eq.~\eqref{eq:H}]. (b) Mean-field phase diagram with normal (NP), chiral superradiant (CSR), and degenerate ferro- and antiferro-superradiant (FSR/AFSR) phases. On the matched line $\beta=\beta_*=1/9$ with $g^2>r_*=0.82$ (thick line), all $2^N$ sign configurations are degenerate; the dashed extension marks $\beta_*$ in the normal phase. Points A and B mark $x=g^2/r_*=1.2$ and $1.4$. (c) Coherent photon displacements $\alpha_n/\sqrt{\eta}$ [Eqs.~\eqref{eq:Xn} and \eqref{eq:alpha}] at A on a 12-site ring for FSR, AFSR, CSR, and a generic configuration with the full ring period. Each arrow starts at site $n$ and shows $\mathrm{Re}\,\alpha_n$ upward and $\mathrm{Im}\,\alpha_n$ to the right, on a common scale; blue and orange arrows have $\tau_n=+1$ and $-1$, and the signs above the sites list $\tau_n$.}
\label{fig:model}
\end{figure*}

We consider a periodic ring of $N$ cavities with a two-level atom at each site [Fig.~\ref{fig:model}(a)]. In units with $\hbar=1$,
\begin{align}
H={}&\omega\sum_n a_n^\dagger a_n-J\sum_n\bigl(e^{-i\theta}a_n^\dagger a_{n+1}+\text{H.c.}\bigr)\nonumber\\
&+\frac{\Delta}{2}\sum_n\sigma_n^z+\lambda\sum_n\sigma_n^x Q_n,\label{eq:H}\\
Q_n={}&\frac{X_n+\beta X_{n+2}}{\sqrt{1+\beta^2}},\qquad X_n=a_n+a_n^\dagger.\label{eq:Q}
\end{align}
Here $a_n$ annihilates a cavity photon, $\omega$ and $\Delta$ are the cavity and atomic transition frequencies, and $\sigma_n^{x,z}$ are Pauli operators for atom $n$, represented as a spin-$1/2$. Atom $n$ couples through $\sigma_n^x$ to the contact quadrature $Q_n$, a linear combination of the field quadratures in cavities $n$ and $n+2$ on the same sublattice. The parameter $\beta$ specifies the ratio of the two contact strengths, whose squared magnitudes sum to $\lambda^2$.

We use the dimensionless parameters
\begin{equation}
j=\frac{2J}{\omega},\qquad g=\frac{2\lambda}{\sqrt{\omega\Delta}},\qquad \eta=\frac{\Delta}{\omega},\label{eq:params}
\end{equation}
with $0<j<1$, $0\le\beta<1$, $\theta=\pi/2$, and $N$ a multiple of 4. The bare photon band $\omega_q=\omega(1-j\sin q)$ is then positive and dispersive. As in the single-site quantum Rabi model, the superradiant transition occurs for $\eta\to\infty$ at fixed $g$ and $j$~\cite{Hwang2015}. For $N$ a multiple of 4, the gauge transformation $a_n\mapsto i^na_n$ preserves the periodic boundary conditions and renders the hopping real. It also rotates the odd-site contact quadratures from $X_n$ to $\pm P_n$, where $P_n=i(a_n^\dagger-a_n)$. The light--matter coupling therefore fixes the hopping phase relative to the contact quadratures. Throughout this work, $\theta$ is defined in the gauge with real contact amplitudes (Appendix~\ref{app:static}).

We evaluate the mean-field energy using coherent photon states with amplitudes $\alpha_n=\langle a_n\rangle$ and the local spin ground states, and define the scaled contact fields $s_n=\langle Q_n\rangle/\sqrt{\eta}$. At $\theta=\pi/2$, the hopping couples each quadrature $X_n$ to the conjugate quadratures $P_{n\pm1}$ of the neighboring cavities. Minimization over the $P$ quadratures yields a static stiffness coupling $X$ quadratures separated by two sites, and the mean-field energy becomes
\begin{align}
\frac{E_\mathrm{MF}[s]}{\Delta}={}&\frac14\sum_q\rho_\beta(q)|s_q|^2-\frac12\sum_n\sqrt{1+g^2s_n^2},\label{eq:EMF}\\
\rho_\beta(q)={}&\frac{h(q)}{|F_\beta(q)|^2},\qquad F_\beta(q)=\frac{1+\beta e^{2iq}}{\sqrt{1+\beta^2}},\label{eq:rho}
\end{align}
where $s_q=N^{-1/2}\sum_ne^{-iqn}s_n$ is the unitary Fourier transform of $s_n$, $h(q)=1-j^2\sin^2q$ is the static photon stiffness, and $F_\beta(q)$ is the form factor of the two contacts. The function $\rho_\beta(q)$ determines the momentum-dependent instability threshold: in the normal phase, the lower excitation energy in each momentum pair $(q,-q)$ vanishes at $g^2=g_c^2(q)=\rho_\beta(q)$ (Appendix~\ref{app:static}). Both $h(q)$ and $|F_\beta(q)|^2$ are linear in $\cos2q$, so their ratio is independent of momentum at
\begin{equation}
\beta_*=\frac{1-\sqrt{1-j^2}}{1+\sqrt{1-j^2}},\qquad \rho_{\beta_*}(q)=r_*\equiv1-\frac{j^2}{2}.\label{eq:match}
\end{equation}
At this contact ratio, all momentum pairs become unstable at the same coupling $g^2=r_*$. Since $\rho_\beta(q)=r_*$ at $\cos2q=0$ for every $\beta$, the matched contact ratio maximizes the normal-phase threshold $\min_q\rho_\beta(q)$ [Fig.~\ref{fig:model}(b)]. A separation of two lattice spacings is the shortest that allows this matching at $\theta=\pi/2$. On the infinite chain, the inverse static stiffness has normalized coefficients $(-\beta_*)^{|\ell|}$ at distance $2\ell$. The form factor $|F_{\beta_*}(q)|^2=h(q)/r_*$ cancels these nonlocal contributions in the static contact response. Extensions to other hopping phases and contact geometries are discussed in Appendix~\ref{app:static}.

At the matched contact ratio, we define $x=g^2/r_*$. Parseval's identity reduces the first term of Eq.~\eqref{eq:EMF} to $r_*\sum_ns_n^2/4$, and the mean-field energy becomes a sum of identical single-site Rabi energies,
\begin{equation}
E_\mathrm{MF}=\Delta\sum_nw(s_n),\qquad w(s)=\frac{r_*s^2}{4}-\frac12\sqrt{1+g^2s^2}.\label{eq:w}
\end{equation}
For $x>1$, each local potential is a double well with minima at
\begin{equation}
s_n=\tau_ns_0,\qquad \tau_n=\pm1,\qquad s_0^2=\frac{x^2-1}{r_*x},\label{eq:s0}
\end{equation}
separated by the barrier
\begin{equation}
\Delta E_b=\Delta B(x),\qquad B(x)=\frac{x+x^{-1}-2}{4}.\label{eq:barrier}
\end{equation}
Each of the $N$ contact fields can independently occupy either minimum, giving $2^N$ degenerate mean-field configurations. The Ising variable $\tau_n$ labels the sign of the contact field, and the corresponding local spin polarization is $\langle\sigma_n^x\rangle=-\tau_n\sqrt{1-x^{-2}}$.

Each sign configuration corresponds to a distinct configuration of photon displacements. Inverting Eq.~\eqref{eq:Q} gives
\begin{equation}
\langle X_n\rangle=\sqrt{\eta(1+\beta_*^2)}\,s_0\sum_{m\ge0}(-\beta_*)^m\tau_{n+2m},\label{eq:Xn}
\end{equation}
with indices taken modulo $N$, and the minimization over the conjugate quadratures gives
\begin{equation}
\langle P_n\rangle=\frac j2\bigl(\langle X_{n-1}\rangle-\langle X_{n+1}\rangle\bigr),\qquad \alpha_n=\frac{\langle X_n\rangle+i\langle P_n\rangle}{2}.\label{eq:alpha}
\end{equation}
Since $\sum_{m\ge1}\beta_*^m<1$ for $\beta_*<1/2$, that is, for $j<2\sqrt2/3$, the $m=0$ term in Eq.~\eqref{eq:Xn} fixes the sign of $\langle X_n\rangle=2\,\mathrm{Re}\,\alpha_n$ to be $\tau_n$. Within this parameter range, we refer to $\tau_n$ as the displacement sign. Figure~\ref{fig:model}(c) shows four of the degenerate configurations. Uniform and staggered signs yield the ferro-superradiant (FSR) and antiferro-superradiant (AFSR) configurations, respectively, with real displacements and $\langle P_n\rangle=0$. The period-four pattern $(++--)$ gives
\begin{equation}
(\alpha_0,\alpha_1,\alpha_2,\alpha_3)\propto(1-ij,1+ij,-1+ij,-1-ij),\label{eq:CSR}
\end{equation}
whose phase winds once every four sites. A generic sign configuration can produce a complex displacement pattern with no spatial period shorter than the ring. In momentum space Eq.~\eqref{eq:alpha} reads $\alpha_q=(1+j\sin q)\langle Q_q\rangle/[2F_{\beta_*}(q)]$, so the coherent photon number is
\begin{equation}
\sum_n|\alpha_n|^2=\eta\sum_qW(q)|s_q|^2,\qquad W(q)=\frac{(1+j\sin q)^2}{4|F_{\beta_*}(q)|^2}.\label{eq:W}
\end{equation}
For fixed $j<1$, $W(q)$ is bounded above and bounded away from zero, so the coherent photon number scales as $N\eta s_0^2$ for every configuration.

Away from the matched contact ratio, $\rho_\beta(q)$ is a monotonic function of $\cos2q$, and its derivative with respect to $\cos2q$ changes sign at $\beta_*$. Its minimum lies at $q=\pm\pi/2$ for $\beta<\beta_*$ and at $q=0,\pi$ for $\beta>\beta_*$, and the critical coupling is
\begin{equation}
g_c^2=\rho_\mathrm{min}=
\begin{cases}
(1+\beta^2)(1-j^2)/(1-\beta)^2, & \beta<\beta_*,\\
r_*, & \beta=\beta_*,\\
(1+\beta^2)/(1+\beta)^2, & \beta>\beta_*.
\end{cases}\label{eq:gc}
\end{equation}
Below $g_c$ the system is in the normal phase (NP). Replacing $\rho_\beta(q)$ by $\rho_\mathrm{min}$ in Eq.~\eqref{eq:EMF} bounds the energy from below by a sum of local potentials. Contact fields with constant magnitude and Fourier components restricted to the minimizing momenta saturate this bound and yield the global minima (Appendix~\ref{app:static}). For $\beta<\beta_*$, these are the four period-four sign patterns related to $(++--)$ by translation. These correspond to the chiral superradiant (CSR) configuration of the quantum Rabi square~\cite{FallasPadilla2022} repeated around the ring. For $\beta>\beta_*$, they are the uniform and staggered configurations of the FSR and AFSR phases, which are degenerate at $\theta=\pi/2$ (Sec.~\ref{sec:quadratic}). All $2^N$ configurations are degenerate on the matched line $\beta=\beta_*$ with $g^2>r_*$, where these phases meet [Fig.~\ref{fig:model}(b)]. The matched line is therefore a multiphase line in the sense of Fisher and Selke~\cite{FisherSelke1980}, with an extensive degeneracy. We use two reference points on this line at $j=0.6$: A ($x=1.2$, $g^2=0.984$) and B ($x=1.4$, $g^2=1.148$).

\section{Harmonic fluctuations and symmetry}\label{sec:quadratic}

\begin{figure*}[t]
\includegraphics[width=\textwidth]{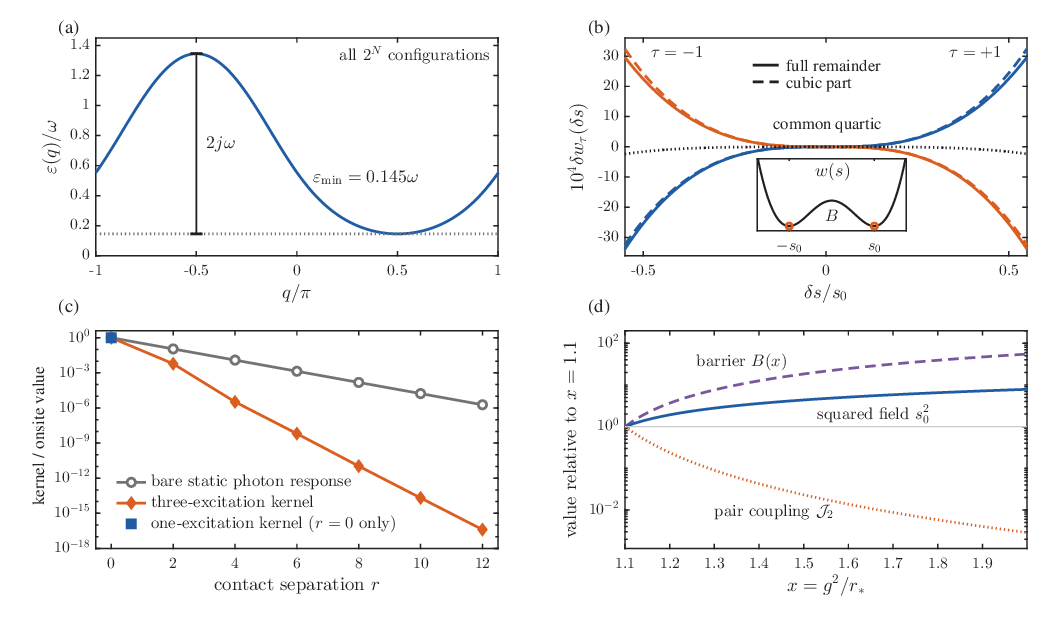}
\caption{Harmonic spectrum and anharmonic selection at A ($j=0.6$, $\theta=\pi/2$, $\beta_*=1/9$, $x=1.2$). (a) Excitation spectrum of Eq.~\eqref{eq:excitation}, common to all $2^N$ configurations, with gap $\varepsilon_\mathrm{min}$ at $q=\pi/2$ and bandwidth $2j\omega$. (b) Anharmonic part of the local potential, $\delta w_\tau(\delta s)=w(\tau s_0+\delta s)-w(s_0)-w''(s_0)\,\delta s^2/2$, for $\tau=\pm1$: full remainder (solid), cubic part (dashed), and common quartic part (dotted). Inset: the double well $w(s)$ with barrier $B(x)$. (c) Infinite-chain real-space kernels normalized to their onsite values, at even separations $r$ in lattice spacings: the one-excitation integral of Eq.~\eqref{eq:local}, nonzero only at $r=0$; the three-excitation integral $\mathrm{Re}\int_0^\infty du\,G_r(u)^3$; and the real-space Fourier coefficients of the bare static photon response $2/[\omega h(q)]$, with normalized magnitudes $\beta_*^{r/2}$. (d) $B(x)$, $s_0^2$, and $\mathcal{J}_2$ along the matched line, relative to their values at $x=1.1$.}
\label{fig:mechanism}
\end{figure*}

To compare quantum fluctuations about the mean-field configurations, we expand the Hamiltonian to quadratic order following the large-$\eta$ expansion of the quantum Rabi model~\cite{Hwang2015,Ye2025}. For an arbitrary configuration $\tau=\{\tau_n\}$, we displace each cavity field by its mean-field value $\alpha_n$, rotate the spin basis to diagonalize the local spin Hamiltonian, and introduce bosons $b_n$ for small spin fluctuations (Appendix~\ref{app:quadratic}). In terms of the contact fluctuations $\phi_n=Q_n-\sqrt{\eta}\,\tau_ns_0$, the quadratic Hamiltonian is
\begin{equation}
H_\tau^{(2)}=H_\mathrm{ph}^{(2)}+\Delta x\sum_nb_n^\dagger b_n+\frac{\lambda}{x}\sum_n\phi_n\bigl(b_n+b_n^\dagger\bigr).\label{eq:H2}
\end{equation}
Here $H_\mathrm{ph}^{(2)}$ is the photon part of Eq.~\eqref{eq:H} with each $a_n$ replaced by $a_n-\alpha_n$. Terms linear in the fluctuations vanish by stationarity. A displacement does not change the quadratic coefficients of the photon Hamiltonian. The local spin splitting $\Delta x$ and the transverse coupling $\lambda/x$ are identical in the two minima, so $H_\tau^{(2)}$ is independent of the displacement signs. The first sign-dependent contribution is proportional to $\tau_n\phi_nb_n^\dagger b_n$ and is cubic in the fluctuation operators. At leading order in the large-$\eta$ expansion, a Schrieffer--Wolff transformation eliminates the spin-flip coupling and yields the effective photon Hamiltonian
\begin{equation}
H_\mathrm{eff}^{(2)}=H_\mathrm{ph}^{(2)}-\frac{\omega r_*}{4x^2}\sum_n\phi_n^2+\mathrm{const.}
\label{eq:SW}
\end{equation}

Bogoliubov diagonalization of Eq.~\eqref{eq:SW} gives the common photon excitation spectrum
\begin{equation}
\begin{split}
\frac{\varepsilon(q)}{\omega}&=e_q-j\sin q,\\
e_q&=\sqrt{j^2\sin^2q+(1-x^{-2})(1-j^2\sin^2q)},
\end{split}\label{eq:excitation}
\end{equation}
for all $2^N$ configurations [Fig.~\ref{fig:mechanism}(a)]. For $x>1$, $\varepsilon(q)$ is a monotonically decreasing function of $\sin q$. Its minimum occurs at $q=\pi/2$, with the positive gap $\varepsilon_\mathrm{min}=\omega\bigl[\sqrt{1-(1-j^2)x^{-2}}-j\bigr]$, while its maximum is $\varepsilon_\mathrm{min}+2j\omega$ at $q=-\pi/2$. The gap is $0.1454\,\omega$ at A. Thus the common spectrum remains dispersive, with bandwidth $2j\omega$, while the spin excitation energies are of order $\Delta$. All configurations also share the same zero-point energy and harmonic correlation functions. The $2^N$ configurations therefore remain degenerate at harmonic order.

A symmetry of the full Hamiltonian constrains the effective interactions between displacement signs. Let $\mathcal{K}$ denote complex conjugation in the photon-number and $\sigma^z$ basis. At $\theta=\pi/2$, the antiunitary operation
\begin{equation}
\mathcal{A}=U_\mathrm{sub}\mathcal{K},\qquad U_\mathrm{sub}:(a_n,\sigma_n^x)\mapsto(-1)^n(a_n,\sigma_n^x),\label{eq:antiunitary}
\end{equation}
leaves Eq.~\eqref{eq:H} invariant. Complex conjugation reverses the hopping phase, and the staggered transformation restores the original hopping term. The light--matter coupling is invariant because each atom couples to two cavities on the same sublattice. The operation acts on the contact fields as $Q_n\mapsto(-1)^nQ_n$ and hence on their signs as $\tau_n\mapsto(-1)^n\tau_n$. Together with global parity, $(a_n,\sigma_n^x)\mapsto-(a_n,\sigma_n^x)$, which maps $\tau_n\mapsto-\tau_n$, this symmetry makes the configuration energy invariant under independent sign reversals on the two sublattices. Its expansion can therefore contain only products with an even number of signs from each sublattice. Pair couplings at odd separation vanish to all orders in $\eta^{-1/2}$, and the leading pair interactions act within each sublattice. The same antiunitary symmetry relates the FSR and AFSR configurations and enforces equality of their configuration energies.

\section{Quantum order by disorder}\label{sec:selection}

Anharmonic terms give the leading configuration-dependent correction to the common quadratic theory. Projecting each spin onto its local ground state at fixed $Q_n$ gives the adiabatic spin potential $V(Q_n)=-\sqrt{\Delta^2+4\lambda^2Q_n^2}/2$. Together with the photon part of Eq.~\eqref{eq:H}, this potential defines the adiabatic Hamiltonian $H_\mathrm{ad}$. The full local potential $w(s)$ has two minima with equal curvature and quartic coefficients but opposite cubic coefficients [Fig.~\ref{fig:mechanism}(b)]. The cubic contribution is
\begin{equation}
V_3=\frac{v_3}{6}\sum_n\tau_n\phi_n^3,\qquad v_3=\frac{3\omega}{2\sqrt{\eta}}\,\frac{r_*^{3/2}\sqrt{x^2-1}}{x^{7/2}}.\label{eq:V3}
\end{equation}
The second-order cubic correction and the first-order quartic correction are both of order $\omega/\eta$, but only the cubic vertex depends on the displacement signs (Appendix~\ref{app:kernel}).

The second-order correction from $V_3$ contains products $\tau_n\tau_m$, with one sign from each vertex. It therefore generates an effective Ising interaction,
\begin{equation}
\begin{split}
H_\mathrm{Ising}[\tau]&=-\sum_{r\ge1}J_r\sum_n\tau_n\tau_{n+r}+\text{const.},\\
J_r&=\frac{\omega}{\eta}\,\mathcal{J}_r(j,x),
\end{split}\label{eq:Epat}
\end{equation}
where the scaling follows from $v_3^2/\omega\propto\omega/\eta$. The dimensionless coefficients $\mathcal{J}_r(j,x)$ depend on the photon hopping and the coupling along the matched line. This Hamiltonian describes the leading configuration-dependent energies of states localized near the mean-field minima. The symmetry established in Sec.~\ref{sec:quadratic} restricts the pair interactions to even separations.

The second-order correction involves intermediate states with either one or three Bogoliubov excitations. With $\mu=1-x^{-2}$ and the dimensionless imaginary time $u=\omega t_E$, where $t_E$ is the Euclidean time, the common harmonic correlation function $G_{n-m}(u)=\langle\phi_n(u)\phi_m(0)\rangle_0$ is, in the infinite-chain limit,
\begin{equation}
G_r(u)=\int_{-\pi}^{\pi}\frac{dq}{2\pi}\,e^{iqr}\,\frac{|F_{\beta_*}(q)|^2}{e_q}\,e^{-(e_q-j\sin q)u},\label{eq:G}
\end{equation}
for $u\ge0$. Normal ordering gives $\phi_n^3={:}\phi_n^3{:}+3G_0(0)\phi_n$. The resulting linear term couples to one-excitation intermediate states and corresponds to a force induced by zero-point fluctuations, $-v_3G_0(0)\tau_n/2$, acting on the contact field. The static response to these forces is
\begin{equation}
\frac{\omega}{2}\chi^{\mathrm{stat}}_{QQ}(r)=\mathrm{Re}\int_0^\infty du\,G_r(u)=\frac{\delta_{r0}}{\mu r_*}.\label{eq:local}
\end{equation}
Its locality follows from the same contact matching that factorizes the mean-field energy [Fig.~\ref{fig:mechanism}(c)]. Including the spin-induced term in Eq.~\eqref{eq:SW}, the static quadratic energy of the contact fluctuations is $(\omega r_*\mu/4)\sum_n\phi_n^2$, which is onsite. The one-excitation contribution is therefore proportional to $\sum_n\tau_n^2=N$ and gives the same energy shift for every configuration. The three-excitation channel involves sums of three excitation energies in its denominators. Its contribution remains nonlocal and gives
\begin{equation}
J_r=\frac{v_3^2}{3\omega}\,\mathrm{Re}\int_0^\infty du\,G_r(u)^3.\label{eq:Jr}
\end{equation}
The three-propagator kernel decays more rapidly with distance than the bare static photon response [Fig.~\ref{fig:mechanism}(c)]. The identity $G_r(u)^*=(-1)^rG_r(u)$ makes the odd-separation couplings vanish, as required by the symmetry of Eq.~\eqref{eq:antiunitary}. The Wick contractions and numerical evaluation are given in Appendix~\ref{app:kernel}.

At point A, the infinite-chain coefficients are
\begin{equation}
\begin{split}
&\mathcal{J}_2=7.283\times10^{-4},\\
&\mathcal{J}_4/\mathcal{J}_2=5.522\times10^{-4},\quad \mathcal{J}_6/\mathcal{J}_2=1.066\times10^{-6}.
\end{split}\label{eq:Jvalues}
\end{equation}
The positive $J_2$ favors equal signs on neighboring sites of each sublattice, while the longer-range couplings are smaller by more than three orders of magnitude. The effective interaction thus consists of two sublattice Ising chains dominated by a ferromagnetic nearest-neighbor coupling $J_2$. The minimum-energy configurations have uniform signs within each sublattice, giving the two FSR and two AFSR configurations. Reversing a consecutive block of signs on one sublattice creates two domain walls with a total energy close to $4J_2$. At A, the energy of two well-separated walls is $4.004J_2$ (Appendix~\ref{app:kernel}).

This selection has no classical counterpart. In the classical phase-space integral, a coordinate-dependent shift of the momenta removes the effective vector potential associated with the $X$--$P$ coupling, as in the Bohr--van Leeuwen theorem. The remaining Gaussian momentum integral is independent of the coordinates, so the contact-field distribution factorizes,
\begin{equation}
P_\mathrm{cl}(Q)\propto\prod_n\exp\Bigl[-\frac{\Delta}{k_{\mathrm{B}}T}\,w\bigl(Q_n/\sqrt{\eta}\bigr)\Bigr],\label{eq:Pcl}
\end{equation}
with the full local nonlinearity retained (Appendix~\ref{app:quadratic}). Since $w$ is even, all sign configurations have equal classical weight at every temperature. In the common harmonic ground state, by contrast, the contact fluctuations are correlated between sites; at A, $\langle\phi_n\phi_{n+2}\rangle_0/\langle\phi_n^2\rangle_0=0.18$. The cubic terms convert these correlations into the configuration-dependent energy through the three-excitation processes of Eq.~\eqref{eq:Jr}, and the resulting selection is therefore purely quantum.

Increasing the atom--photon coupling strengthens the local displacements while reducing the induced interaction. Along the matched line, increasing $x$ from 1.1 to 2 raises $s_0^2$ by nearly a factor of eight and the barrier by a factor of 55, while $\mathcal{J}_2$ decreases by a factor of approximately 347 [Fig.~\ref{fig:mechanism}(d)]. The excitation energies in the virtual three-excitation processes increase with $x$; for $x>\sqrt{7/5}$, the squared cubic vertex $v_3^2\propto(x^2-1)/x^7$ also decreases. At fixed $j$ and $x$, the barrier and pair coupling satisfy
\begin{equation}
\frac{\Delta E_b}{J_2}=\frac{B(x)}{\mathcal{J}_2(j,x)}\,\eta^2,\label{eq:ratio}
\end{equation}
while $\varepsilon_\mathrm{min}/J_2$ grows linearly with $\eta$. At A and $\eta=256$, $\Delta E_b=2.133\,\omega$ and $J_2=2.845\times10^{-6}\omega$, giving $\Delta E_b/J_2=7.50\times10^5$. The interaction that selects the displacement signs is therefore well below both the excitation gap and the local barrier.

\section{Four-site calculations}\label{sec:checks}

\begin{figure*}[t]
\includegraphics[width=\textwidth]{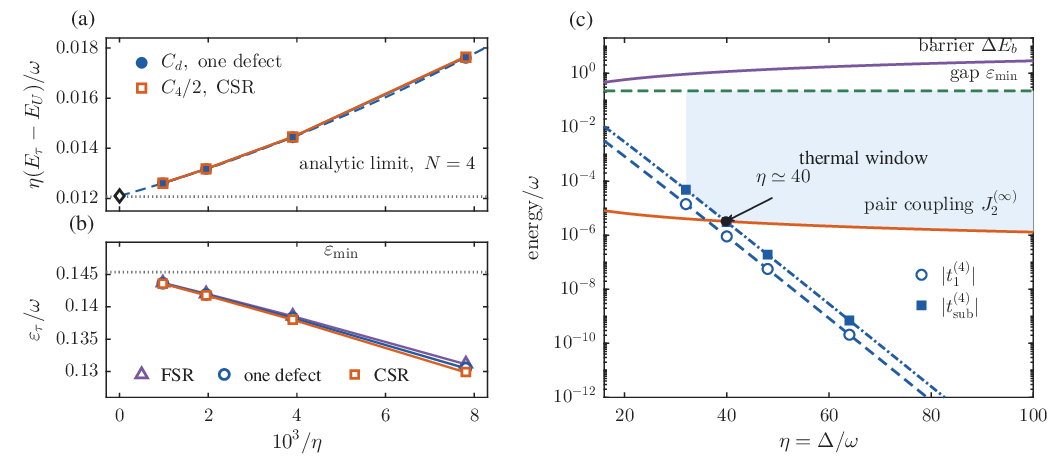}
\caption{Four-site full-spin calculations at $j=0.6$, $\theta=\pi/2$, and $\beta_*=1/9$; panels (a) and (b) at A, panel (c) at B. (a) Scaled energy differences $C_d$ and $C_4/2$ relative to FSR [Eq.~\eqref{eq:Ctau}]. Symbols are cutoff-extrapolated full-spin results and solid lines adiabatic results. The dashed line is the quadratic full-spin fit, extrapolated to the open diamond at $\eta^{-1}=0$; the fits for $C_d$ and $C_4/2$ coincide on this scale. The dotted line is the four-site analytic limit. (b) Intrawell excitation gaps at $n_c=18$. Symbols are full-spin results and solid lines are adiabatic results; the dotted line is $\varepsilon_\mathrm{min}$. (c) Barrier, gap, infinite-chain coupling $J_2^{(\infty)}$, and four-site tunneling amplitudes for single (circles) and whole-sublattice (squares) flips, with exponential fits (dashed). The dot marks where the whole-sublattice fit crosses $J_2^{(\infty)}$. For $\eta\ge32$, shading spans the interval between $\max(J_2^{(\infty)},|t_1^{(4)}|,|t_\mathrm{sub}^{(4)}|)$ and $\varepsilon_\mathrm{min}$.}
\label{fig:checks}
\end{figure*}

We examine the configuration energies and tunneling amplitudes on a four-site ring, retaining all sixteen spin states (full-spin calculations; Appendix~\ref{app:fullspin}). For each configuration, we use an extended coherent-state (displaced Fock) basis~\cite{Chen2008,Chen2012} centered on the mean-field displacements, with a cutoff $n_c$ on the total number of displaced photons. Within a range of cutoffs, the lowest Ritz values form a plateau and the corresponding states remain localized near the selected minimum. These values give the intrawell configuration energies described by Eq.~\eqref{eq:Epat}. Configuration energies and interwell tunneling are distinct contributions to the finite-size spectrum~\cite{Khatua2026}. We denote the FSR configuration $(++++)$, the one-defect configuration $(+++-)$, and the CSR configuration $(++--)$ by $U$, $d$, and $4$, respectively, and define the scaled energy differences
\begin{equation}
C_\tau(\eta)=\frac{\eta}{\omega}\bigl[E_\tau(\eta)-E_U(\eta)\bigr].\label{eq:Ctau}
\end{equation}

For a four-site ring, the momentum integral in Eq.~\eqref{eq:G} is replaced by a sum over the four allowed momenta. Including contributions from paths that wind around the ring gives the analytic limits $C_d^{(4)}=0.01207$ and $C_4^{(4)}=2C_d^{(4)}$. The one-defect configuration changes the relative signs within one sublattice, while the CSR configuration changes them within both. Their energy differences therefore probe the sublattice pair interactions in Eq.~\eqref{eq:Epat}. Figure~\ref{fig:checks}(a) shows the results at A for $\eta=128$--$1024$. The full-spin and adiabatic values of $C_\tau(\eta)$ agree within $0.2\%$ throughout this range, showing that the adiabatic projection reproduces the configuration energy differences. Cutoff extrapolation followed by a quadratic fit in $\eta^{-1}$ gives intercepts of $0.01210$ and $0.02420$, within $0.24\%$ of the analytic limits. Their ratio is close to two, as predicted by the leading pair interaction. The cutoff dependence and sensitivity to the fitting range and polynomial order are given in Appendix~\ref{app:fullspin}.

The intrawell excitation gaps provide a further comparison with the common quadratic theory [Fig.~\ref{fig:checks}(b)]. For the FSR, one-defect, and CSR configurations, the gaps converge to the same harmonic limit $\varepsilon_\mathrm{min}$ as $\eta^{-1}\to0$. The allowed momenta of the four-site ring include $q=\pi/2$, where Eq.~\eqref{eq:excitation} is minimal, so this limit equals the infinite-chain gap. The full-spin and adiabatic gaps nearly coincide over the whole range (Appendix~\ref{app:fullspin}).

Tunneling between configurations introduces a second low-energy scale. With $Q=\sqrt{\eta}\,s$, the Euclidean action of $H_\mathrm{ad}$ is proportional to $\eta$, giving tunneling amplitudes that decrease exponentially with $\eta$, while $J_2$ decreases as $\eta^{-1}$. Figure~\ref{fig:checks}(c) shows full-spin matrix elements between orthogonalized localized states at B, where $\varepsilon_\mathrm{min}=0.2207\,\omega$, for $\eta=32$--$64$. Both the single-flip amplitude $t_1^{(4)}$ and the whole-sublattice amplitude $t_\mathrm{sub}^{(4)}$ exhibit exponential decay. The fitted decay coefficients agree with the corresponding saddle-point actions per unit $\eta$ within $0.6\%$ (Appendix~\ref{app:fullspin}). At B, the two flips have nearly equal saddle-point actions, and the larger whole-sublattice amplitude reflects its larger prefactor (Appendix~\ref{app:fullspin}). For the four-site ring, $|t_\mathrm{sub}^{(4)}|$ falls below the infinite-chain coupling $J_2^{(\infty)}$ near $\eta=40$, and its ratio to $J_2^{(\infty)}$ is $3.4\times10^{-4}$ at $\eta=64$. These results illustrate the separation between exponentially small tunneling amplitudes and the algebraically small interaction that selects the displacement signs.

\section{Thermal correlations and photon occupation}\label{sec:consequences}

\begin{figure*}[t]
\includegraphics[width=\textwidth]{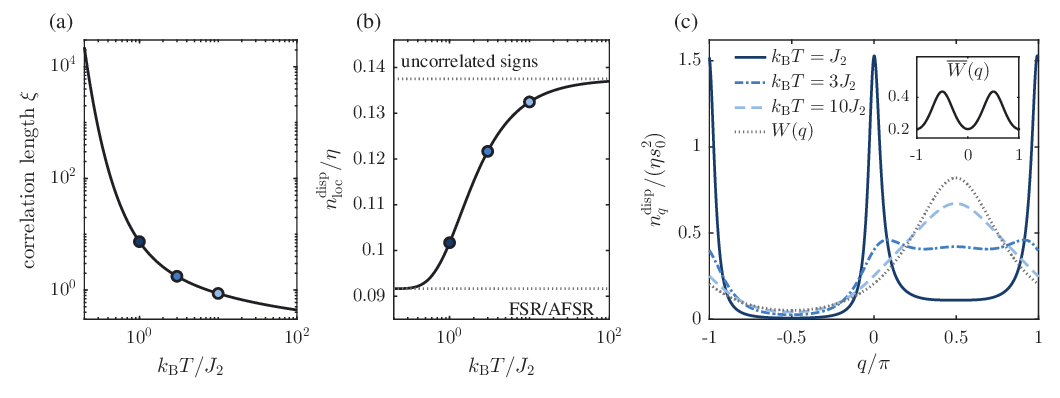}
\caption{Thermal sign correlations and photon occupation at A ($j=0.6$, $x=1.2$, $\theta=\pi/2$, $\beta_*=1/9$) in the thermodynamic limit, for the two sublattice Ising chains with couplings $J_2$--$J_8$. (a) Correlation length $\xi$ in physical lattice spacings. (b) Displacement contribution to the photon number per site, $n_\mathrm{loc}^\mathrm{disp}/\eta$; the dotted lines mark the FSR/AFSR value $0.0917$ and the uncorrelated-sign limit $0.1375$. Colored dots in (a) and (b) mark the temperatures used in (c). (c) Photon momentum distribution $n_q^\mathrm{disp}/(\eta s_0^2)=W(q)S_\tau(q)$ over the full Brillouin zone at $k_{\mathrm{B}}T/J_2=1$, 3, and 10; all profiles are symmetric about $q=\pi/2$. The dotted curve is the uncorrelated-sign limit $W(q)$ of Eq.~\eqref{eq:W}. Inset: symmetrized weight $\overline W(q)=[W(q)+W(-q)]/2$, which is smallest at $q=0$ and $\pi$.}
\label{fig:thermal}
\end{figure*}

When tunneling and thermal excitations within the wells are negligible, the leading configuration energies define a classical Ising ensemble for the displacement signs. The separation of energy scales permits a weakly correlated regime,
\begin{equation}
\max\Bigl(|t|,\sum_m|J_{2m}|\Bigr)\ll k_{\mathrm{B}}T\ll\min(\varepsilon_\mathrm{min},\Delta E_b).\label{eq:window}
\end{equation}
Here $t$ is the relevant local tunneling amplitude [Fig.~\ref{fig:checks}(c)]. The lower bound places both the sign interactions and tunneling below the thermal scale. The upper bound suppresses intrawell excitations and thermal activation over the barrier, keeping each contact field near a local minimum; thermal corrections to the couplings are then exponentially small (Appendix~\ref{app:thermal}). Once tunneling is negligible, the ratio of the upper and lower scales grows as $\varepsilon_\mathrm{min}/J_2\propto\eta$. At A and $\eta=256$, the interval $10J_2=2.85\times10^{-5}\omega<k_{\mathrm{B}}T<0.1\varepsilon_\mathrm{min}=1.45\times10^{-2}\omega$ has an upper bound 511 times its lower bound.

At leading order, Eq.~\eqref{eq:Epat} describes two independent classical Ising chains, one on each sublattice. To follow the crossover into the weakly correlated regime, we also evaluate their correlations at temperatures of order $J_2/k_{\mathrm{B}}$. The chains have no long-range order at positive temperature, and their sign correlations decay exponentially with distance. A transfer-matrix calculation retaining $J_2$--$J_8$ (Appendix~\ref{app:thermal}) gives a correlation length that decreases from $\xi=7.36$ physical lattice spacings at $k_{\mathrm{B}}T=J_2$ to $0.87$ at $k_{\mathrm{B}}T=10J_2$ [Fig.~\ref{fig:thermal}(a)]. Since $J_2$ dominates, these values are close to the nearest-neighbor result $\xi=2/\ln\coth(J_2/k_{\mathrm{B}}T)$. The displacement signs are nearly uncorrelated at $k_{\mathrm{B}}T=10J_2$.

The sign correlations determine the photon momentum distribution through Eq.~\eqref{eq:W},
\begin{align}
n_q^\mathrm{disp}&=\eta s_0^2\,W(q)\,S_\tau(q),\label{eq:nq}\\
S_\tau(q)&=\Bigl\langle\Bigl|\frac{1}{\sqrt{N}}\sum_ne^{-iqn}\tau_n\Bigr|^2\Bigr\rangle_T.\label{eq:Sq}
\end{align}
The weight $W(q)$ is fixed by the contact and hopping geometry, while $S_\tau(q)$ describes the thermal sign correlations. The absence of cross-sublattice correlations gives $S_\tau(q+\pi)=S_\tau(q)$. At low temperatures, the structure factor has equal peaks at $q=0$ and $\pi$, reflecting FSR and AFSR correlations, with widths of order $\xi^{-1}$. In the photon distribution, these correlation peaks appear slightly displaced toward $q=\pi/2$ by the slope of $W(q)$ [Fig.~\ref{fig:thermal}(c)]. With increasing temperature, the correlation peaks broaden and decrease in height, while $S_\tau(q)$ approaches unity. The photon distribution approaches $\eta s_0^2W(q)$, whose maximum is at $q=\pi/2$ (Appendix~\ref{app:thermal}). Since $S_\tau(q)$ is even, the asymmetry $n_q^\mathrm{disp}/n_{-q}^\mathrm{disp}=[(1+j\sin q)/(1-j\sin q)]^2$ is set by $W(q)$ alone and does not depend on temperature. Thermal averaging over the configurations gives $\langle a_q\rangle=0$, while the total displacement contribution to the photon number remains of order $N\eta s_0^2$.

The local photon number provides a direct signature of this change in correlations. The displacement contribution to $\langle Q_n^2\rangle$ is $\eta s_0^2$, independent of the signs, whereas the displacement contribution to the photon number per site, $n_\mathrm{loc}^\mathrm{disp}=N^{-1}\sum_qn_q^\mathrm{disp}$, depends on their correlations. It gives the leading contribution to the total photon number $n_\mathrm{loc}=N^{-1}\sum_n\langle a_n^\dagger a_n\rangle$: at fixed $x>1$, $n_\mathrm{loc}=n_\mathrm{loc}^\mathrm{disp}+O(1)$ as $\eta\to\infty$. Fluctuations within each well contribute to the $O(1)$ correction. Since $S_\tau(q)$ is even, the momentum sum is weighted by $\overline W(q)=[W(q)+W(-q)]/2$, which is minimal at $q=0$ and $\pi$ [inset of Fig.~\ref{fig:thermal}(c)]. The FSR and AFSR configurations therefore minimize the coherent photon number among all sign configurations. At A, their coherent photon number is $0.0917\eta$ per site, compared with $0.1948\eta$ for CSR and $0.1375\eta$ for uncorrelated signs. As $k_{\mathrm{B}}T$ increases from $J_2$ to $10J_2$, $n_\mathrm{loc}^\mathrm{disp}$ increases from $0.1017\eta$ to $0.1325\eta$ [Fig.~\ref{fig:thermal}(b)]. Thus the local photon number increases as the sign correlations weaken, while remaining of order $\eta$. Its approach to the uncorrelated-sign limit provides a photon-number signature of the loss of sign correlations.

The results above assume exact contact matching. A uniform deviation $\delta\beta=\beta-\beta_*$ introduces mean-field interactions with the same even-separation structure,
\begin{equation}
\widetilde J_{2m}=J_{2m}+\frac{\Delta s_0^2r_*}{2}(-\beta_*)^{m-1}\delta\beta+O(\Delta\,\delta\beta^2).\label{eq:Jtilde}
\end{equation}
At fixed $j$ and $x$, this classical contribution scales as $\omega\eta\,\delta\beta$, while the quantum interaction scales as $\omega/\eta$. The mismatch-induced contribution favors FSR and AFSR configurations for positive $\delta\beta$ and CSR for negative $\delta\beta$, consistent with the mean-field phase diagram in Fig.~\ref{fig:model}(b). At A, $\widetilde J_2/\omega\simeq\mathcal{J}_2/\eta+0.1833\,\eta\,\delta\beta$, so the quantum term dominates for
\begin{equation}
|\delta\beta|\lesssim\frac{\mathcal{J}_2}{0.1833\,\eta^2},\label{eq:tolQ}
\end{equation}
corresponding to $6.06\times10^{-8}$ at $\eta=256$. A temperature interval with weak sign correlations requires the total sign coupling to remain well below the excitation gap, giving a mismatch tolerance that is parametrically larger by a factor of order $\eta$. Using the upper estimate $K_\mathrm{est}=\sum_m|J_{2m}|+\Delta s_0^2r_*|\delta\beta|/[2(1-\beta_*)]$, we impose the margins $10K_\mathrm{est}<k_{\mathrm{B}}T<0.1\varepsilon_\mathrm{min}$. At A and $\eta=256$, the ratio of the upper to the lower bound decreases from 511 at matching to about 26 at $|\delta\beta|=10^{-6}$. The estimated interval closes at $|\delta\beta|\simeq2.75\times10^{-5}$ for these factor-of-ten margins. Appendix~\ref{app:mismatch} gives the full dependence, the shift of the FSR/AFSR--CSR boundary, and the effects of a hopping-phase offset and spatially varying contacts.

\section{Conclusions}\label{sec:conclusion}

Contact matching in a giant-atom quantum Rabi ring reduces the mean-field energy to a sum of independent local Rabi potentials, giving $2^N$ degenerate superradiant configurations. All configurations share the same quadratic Hamiltonian, dispersive excitation spectrum, and zero-point energy. At large but finite $\eta$, the opposite cubic terms of the two wells generate a same-sublattice Ising interaction of order $\omega/\eta$ through virtual three-excitation processes, selecting ferro- and antiferro-superradiant configurations. The classical contact-field distribution factorizes, so the selection is purely quantum. The induced interaction is much smaller than the excitation gap, allowing macroscopic photon occupation to coexist with short-range sign correlations. As the correlations weaken with increasing temperature, the local photon number rises toward its value for uncorrelated signs while remaining of order $\eta$ per site.

Quantum selection and the weakly correlated thermal regime differ in their sensitivity to contact mismatch. At fixed $j$ and $x$, the tolerance to uniform contact mismatch scales as $\eta^{-2}$ for quantum-dominated selection and as $\eta^{-1}$ for the temperature interval. At A and $\eta=256$, the quantum and mismatch-induced couplings become comparable at $|\delta\beta|\simeq6\times10^{-8}$, while the estimated temperature interval closes at $|\delta\beta|\simeq2.75\times10^{-5}$ with the margins used in Sec.~\ref{sec:consequences}. More generally, contact matching requires the static photon stiffness, obtained by minimizing over the conjugate quadratures, to be proportional to $C^\mathsf{T}C$, where $C$ is the contact matrix (Appendix~\ref{app:static}). In a microscopic implementation, this stiffness must include the additional quadratic terms required by gauge consistency~\cite{Nataf2010,Viehmann2011,DeBernardis2018}. Relevant ingredients have been realized separately: multipoint coupling of superconducting giant atoms~\cite{Kannan2020}, photon hopping in a synthetic magnetic field~\cite{Roushan2017}, analog simulation of ultrastrong and deep-strong Rabi coupling~\cite{Braumuller2017,Langford2017,Lv2018}, and a trapped-ion realization of the Rabi--Hubbard model~\cite{Mei2022}.

\begin{acknowledgments}
This work was supported by the National Natural Science Foundation of China (Grant Nos.~11804240 to S.H. and 12305032 to L.-W.D.) and the Zhejiang Provincial Natural Science Foundation of China (Grant No.~LQ23A050003 to L.-W.D.). ChatGPT (OpenAI) was used for language editing, internal consistency checks, and code packaging and optimization. The authors take full responsibility for the content of this article.
\end{acknowledgments}

\section*{Data availability}
The data and code that support the findings of this article are openly available on Zenodo~\cite{GiantAtomZenodo}.

\appendix

\section{Mean-field reduction and global minima}\label{app:static}

\textit{Static reduction and phase convention.}
With $a_n=N^{-1/2}\sum_qe^{iqn}a_q$, the photon Hamiltonian is $H_\mathrm{ph}=\sum_q\omega_qa_q^\dagger a_q$, where $\omega_q=\omega[1-j\cos(q-\theta)]$. Real contact amplitudes $c_\ell$ at relative positions $d_\ell$ give
\begin{equation}
Q_q=F(q)X_q,\qquad F(q)=\sum_\ell c_\ell e^{iqd_\ell},\qquad \sum_\ell c_\ell^2=1,\label{eq:Fgeneral}
\end{equation}
with $X_q=a_q+a_{-q}^\dagger$. Equation~\eqref{eq:Q} corresponds to $c_\ell=(1,\beta)/\sqrt{1+\beta^2}$ and $d_\ell=(0,2)$. At fixed $X_q$, minimization over the conjugate quadratures in each momentum pair $(q,-q)$ yields the static photon energy
\begin{equation}
\begin{split}
E_\mathrm{ph}^\mathrm{st}&=\frac{\omega}{4}\sum_{q:F(q)\neq0}\frac{h(q)}{|F(q)|^2}|Q_q|^2,\\
h(q)&=\frac{2\omega_q\omega_{-q}}{\omega(\omega_q+\omega_{-q})}.
\end{split}\label{eq:Est}
\end{equation}
Modes with $F(q)=0$ do not couple to the atoms and are minimized independently. Adding the local spin ground-state energy $-\Delta\sqrt{1+g^2s_n^2}/2$ gives Eq.~\eqref{eq:EMF}. At $\theta=\pi/2$, the real-space minimization gives $\langle P_n\rangle=(j/2)(\langle X_{n-1}\rangle-\langle X_{n+1}\rangle)$, as used in Eq.~\eqref{eq:alpha}.

The transformation $a_n\mapsto e^{in\chi}a_n$ shifts $\theta\mapsto\theta-\chi$ and rotates $X_n$ to $\cos(n\chi)X_n-\sin(n\chi)P_n$. For $\chi=\pi/2$ and $N$ a multiple of 4, it preserves the periodic boundary conditions and renders the hopping real. The odd-site contact quadratures are simultaneously rotated to $\pm P_n$, changing the quadratures in the light--matter interaction. Thus $\theta$ is defined relative to the contact quadratures; throughout this work, we use the gauge in which the atoms couple to $X$ quadratures with real amplitudes $c_\ell$.

\begin{figure}[tbp]
\includegraphics[width=\columnwidth]{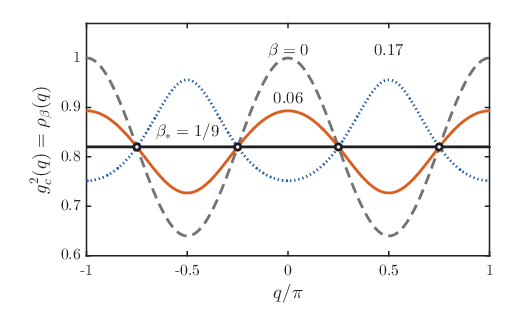}
\caption{Normal-phase instability threshold $g_c^2(q)=\rho_\beta(q)$ at $j=0.6$ and $\theta=\pi/2$ for $\beta=0$, $0.06$, $\beta_*=1/9$, and $0.17$. All curves pass through $g_c^2=r_*=0.82$ at $q=\pm\pi/4$ and $\pm3\pi/4$ (open circles), where $\cos2q=0$. At $\beta_*$, all momentum pairs have the same threshold $g_c^2=r_*$.}
\label{fig:gcq}
\end{figure}

\textit{Contact matching and normal-phase instability.}
For distance-two contacts at $\theta=\pi/2$, $h(q)=1-j^2/2+(j^2/2)\cos2q$ and $|F_\beta(q)|^2=1+2\beta\cos2q/(1+\beta^2)$. Matching the constant and $\cos2q$ coefficients gives $j^2(1+\beta_*)^2=4\beta_*$, yielding Eq.~\eqref{eq:match}. At $\cos2q=0$, $h(q)=r_*$ and $|F_\beta(q)|^2=1$, so the curves $\rho_\beta(q)$ intersect at $r_*$ for $q=\pm\pi/4,\pm3\pi/4$ (Fig.~\ref{fig:gcq}). On the infinite chain,
\begin{equation}
\frac{1}{|1+\beta_*e^{2iq}|^2}=\frac{1}{1-\beta_*^2}\sum_{\ell\in\mathbb Z}(-\beta_*)^{|\ell|}e^{2i\ell q}.
\end{equation}
The bare static response $\chi^\mathrm{ph}_{XX}(q,0)=2/[\omega h(q)]$ therefore has real-space coefficients proportional to $(-\beta_*)^{|\ell|}$ at separation $2\ell$. At matching, the contact response is $|F_{\beta_*}(q)|^2\chi^\mathrm{ph}_{XX}(q,0)=2/(\omega r_*)$, independent of momentum, so all nonlocal coefficients vanish. The form factor has no zeros, with minimum magnitude $(1-\beta_*)/\sqrt{1+\beta_*^2}$.

In the limit $\eta\to\infty$, projection onto the lower spin state gives $H_\mathrm{NP}=H_\mathrm{ph}-(\omega g^2/4)\sum_nQ_n^2+\mathrm{const}$. At $\theta=\pi/2$, Bogoliubov diagonalization in each momentum pair yields
\begin{equation}
\frac{\varepsilon_\mathrm{NP}(\pm q)}{\omega}=\sqrt{1-g^2|F_\beta(q)|^2}\mp j\sin q.\label{eq:normalgap}
\end{equation}
The product of the two excitation energies is $\omega^2[h(q)-g^2|F_\beta(q)|^2]$, which vanishes at $g^2=g_c^2(q)=\rho_\beta(q)$. The normal phase is stable for $g^2<\min_q\rho_\beta(q)$. At matching and at the transition, $\varepsilon_\mathrm{NP}(q)/\omega=|j\sin q|-j\sin q$. Each pair with $0<q<\pi$ has one zero-energy excitation and a partner with energy $2j\omega\sin q$. Together with the zero-energy modes at $q=0$ and $\pi$, this gives $N/2+1$ zero-energy excitations at the common threshold.

\textit{Global mean-field minima.}
The derivative
\begin{equation}
\frac{\partial\rho_\beta}{\partial\cos2q}=\frac{(1+\beta^2)\bigl[j^2(1+\beta)^2-4\beta\bigr]}{2(1+\beta^2+2\beta\cos2q)^2}
\end{equation}
has a fixed sign throughout the Brillouin zone. Consequently, $\rho_\beta(q)$ is minimized at $q=\pm\pi/2$ for $\beta<\beta_*$ and at $q=0,\pi$ for $\beta>\beta_*$. Parseval's identity gives the bounds
\begin{align}
\frac{E_\mathrm{MF}[s]}{\Delta}&\ge\sum_n\Bigl[\frac{\rho_\mathrm{min}}{4}s_n^2-\frac12\sqrt{1+g^2s_n^2}\Bigr]\nonumber\\
&\ge N\min_s\Bigl[\frac{\rho_\mathrm{min}}{4}s^2-\frac12\sqrt{1+g^2s^2}\Bigr].
\end{align}
Writing $y=g^2/\rho_\mathrm{min}$, the local minimum is at $s=0$ for $y\le1$, giving $E_\mathrm{MF}=-N\Delta/2$. For $y>1$, it occurs at $s^2=(y^2-1)/(\rho_\mathrm{min}y)$, with minimum energy $-N\Delta(y+y^{-1})/4$. Both inequalities are saturated by contact fields of constant magnitude whose Fourier components are restricted to the minimizing momenta. Support at $q=\pm\pi/2$ gives $s_{n+2}=-s_n$, yielding the four configurations related to $(++--)$ by translation. Support at $q=0,\pi$ gives $s_{n+2}=s_n$, yielding the uniform and staggered configurations. At matching, all displacement-sign configurations saturate both bounds.

\textit{Photon displacements and occupation bounds.}
The Neumann series of $(I+\beta_*\hat S^2)^{-1}$, where $\hat S$ is the one-site shift, converges for $\beta_*<1$. With indices interpreted modulo $N$, it gives the exact inverse of Eq.~\eqref{eq:Q} and hence Eq.~\eqref{eq:Xn}. The sign criterion $\beta_*<1/2$ in Sec.~\ref{sec:model} is sufficient for every ring size; the bound becomes sharp as $N\to\infty$. To see this, set $\tau_n=+1$ and $\tau_{n+2m}=(-1)^{m+1}$ for $1\le m<N/2$ on one sublattice. Summing the periodic Neumann series gives $\tau_n\langle X_n\rangle\propto1-\sum_{m=1}^{N/2-1}\beta_*^m$, with a positive proportionality factor. For any fixed $\beta_*>1/2$, this expression is negative for sufficiently large $N$.

Since the contact fields are real, $|s_q|^2$ is even in $q$, so the total coherent photon number in Eq.~\eqref{eq:W} depends on $\overline W(q)=[W(q)+W(-q)]/2$. At matching,
\begin{equation}
\overline W(q)=\frac{r_*}{4}\,\frac{1+j^2\sin^2q}{1-j^2\sin^2q}.
\end{equation}
Using $\sum_q|s_q|^2=Ns_0^2$ gives the sharp bounds
\begin{equation}
\frac{N\eta s_0^2r_*}{4}\le\sum_n|\alpha_n|^2\le\frac{N\eta s_0^2r_*}{4}\,\frac{1+j^2}{1-j^2}.
\end{equation}
The lower bound is attained by FSR and AFSR configurations, whose Fourier components lie at $q=0,\pi$, and the upper bound by CSR configurations, with support at $q=\pm\pi/2$. At A, these bounds are $0.0917N\eta$ and $0.1948N\eta$, respectively. Thus every configuration has a coherent photon number of order $N\eta s_0^2$ for fixed $j<1$.

\textit{Matrix matching and spectral factorization.}
Let $K_H$ be the real symmetric, positive-definite static photon stiffness after minimization over the conjugate quadratures, and let $C$ be a real invertible contact matrix. With $Q=CX$ and static energy $\omega X^\mathsf{T}K_HX/4$, the contact-coordinate energy is $(\omega/4)Q^\mathsf{T}C^{-\mathsf{T}}K_HC^{-1}Q$. It is a sum of identical onsite quadratic terms if and only if
\begin{equation}
K_H=r\,C^\mathsf{T}C,\label{eq:matrix}
\end{equation}
with $r>0$. At $\theta=0$, for example, $h(q)=1-j\cos q$ is matched by opposite-sign nearest-neighbor contacts with $j=2\beta/(1+\beta^2)$. At $\theta=\pi/2$, the antiunitary symmetry in Eq.~\eqref{eq:antiunitary} additionally enforces equality of the FSR and AFSR configuration energies.

For a scalar, translationally invariant one-dimensional stiffness, strict positivity and finite range make $h(q)$ a strictly positive trigonometric polynomial. The Fej\'er--Riesz theorem gives a finite polynomial factor~\cite{Sayed2001,Ephremidze2009}. For real, even $h(q)$, the factor can be chosen with real coefficients. Normalizing these coefficients gives finitely many real contact amplitudes and a positive constant $r$ satisfying $h(q)=r|F(q)|^2$. This construction applies to the finite-range stiffnesses at $\theta=0$ and $\pi/2$, with the required number of contacts determined by the polynomial factor. In higher dimensions, a single scalar polynomial factor is not guaranteed; strictly positive trigonometric polynomials instead admit finite sums of squared polynomial moduli~\cite{Dritschel2004}.

\section{Harmonic theory and classical weights}\label{app:quadratic}

\textit{Quadratic Hamiltonian.}
For a mean-field configuration $\tau$, we write $\bar Q_n=\sqrt{\eta}\,\tau_ns_0$ and $Q_n=\bar Q_n+\phi_n$. Since $\lambda\bar Q_n=\tau_n(\Delta/2)\sqrt{x^2-1}$, the local spin Hamiltonian is
\begin{equation}
\frac{\Delta}{2}\sigma_n^z+\lambda\bar Q_n\sigma_n^x=\frac{\Delta x}{2}\bigl(\cos\vartheta\,\sigma_n^z+\tau_n\sin\vartheta\,\sigma_n^x\bigr),
\end{equation}
with $\cos\vartheta=x^{-1}$. We define the rotated Pauli operators by $\tilde\sigma_n^\alpha=R_n\sigma_n^\alpha R_n^\dagger$, where $R_n=e^{-i\tau_n\vartheta\sigma_n^y/2}$. The local Hamiltonian is then $(\Delta x/2)\tilde\sigma_n^z$, with the same splitting $\Delta x$ in both minima. The fluctuation coupling takes the form
\begin{equation}
\lambda\phi_n\sigma_n^x=\lambda\phi_n\Bigl(\frac{\tilde\sigma_n^x}{x}+\tau_n\frac{\sqrt{x^2-1}}{x}\tilde\sigma_n^z\Bigr).
\end{equation}
The local ground state has $\tilde\sigma_n^z=-1$, giving $\langle\sigma_n^x\rangle=-\tau_n\sqrt{1-x^{-2}}$, as used in Sec.~\ref{sec:model}. Introducing bosons through $\tilde\sigma_n^z=-1+2b_n^\dagger b_n$ and $\tilde\sigma_n^x=b_n+b_n^\dagger+O(b^3)$, the linear term $-\tau_n\lambda\sqrt{1-x^{-2}}\,\phi_n$ cancels the linear photon contribution by stationarity. The remaining quadratic terms give Eq.~\eqref{eq:H2}. The transverse coupling $\lambda/x$ is identical in the two minima, and a displacement leaves the quadratic photon coefficients unchanged. The first sign-dependent contribution is $2\tau_n\lambda\sqrt{1-x^{-2}}\,\phi_nb_n^\dagger b_n$, which is cubic in the fluctuation operators.

\textit{Spin projection and retarded response.}
The generator $S=[\lambda/(\Delta x^2)]\sum_n\phi_n(b_n^\dagger-b_n)$ eliminates the coupling $(\lambda/x)\phi_n(b_n+b_n^\dagger)$ in $e^SH_\tau^{(2)}e^{-S}$ at leading order in the large-$\eta$ expansion. Projection onto the bosonic vacuum gives $-\lambda^2\sum_n\phi_n^2/(\Delta x^3)=-(\omega r_*/4x^2)\sum_n\phi_n^2$, as in Eq.~\eqref{eq:SW}. Integrating out the spin bosons within the quadratic theory gives the retarded inverse contact propagator
\begin{equation}
\Pi^R(q,\Omega)=\frac{(\omega_q-\Omega)(\omega_{-q}+\Omega)}{(\omega_q+\omega_{-q})|F(q)|^2}-\frac{2(\lambda/x)^2\Delta x}{(\Delta x)^2-\Omega^2},\label{eq:PiR}
\end{equation}
with $\chi^R_{QQ}=(\Pi^R)^{-1}$. The prescription $\Omega\to\Omega+i0^+$ is understood. This response is common to all configurations. At matching and $\Omega=0$, $\Pi^R=\omega r_*\mu/2$, where $\mu=1-x^{-2}$, reproducing Eq.~\eqref{eq:local}. The collective excitation frequencies follow from the zeros of $\Pi^R$. At matching, they satisfy
\begin{equation}
\bigl[1-(j\sin q+\nu)^2\bigr]\Bigl(x^2-\frac{\nu^2}{\eta^2}\Bigr)=1-j^2\sin^2q,\qquad \nu=\frac{\Omega}{\omega}.
\end{equation}
In the limit $\eta\to\infty$, the low-frequency branch reduces to Eq.~\eqref{eq:excitation}, while the high-frequency branch lies near $\Omega=\Delta x$.

\textit{Contact coordinates and kinetic energy.}
Let $\hat S$ denote the periodic one-site shift, $(\hat SX)_n=X_{n+1}$, and define $P_n=i(a_n^\dagger-a_n)$, so that $[X_n,P_m]=2i\delta_{nm}$. With $\hat B=(j/2)(\hat S-\hat S^\mathsf{T})$, the photon Hamiltonian at $\theta=\pi/2$ is
\begin{equation}
\frac{H_\mathrm{ph}}{\omega}=\frac14\bigl(P+\hat BX\bigr)^\mathsf{T}\bigl(P+\hat BX\bigr)+\frac14X^\mathsf{T}\bigl(I+\hat B^2\bigr)X-\frac N2.
\end{equation}
Minimization over $P$ yields $\langle P\rangle=-\hat B\langle X\rangle$, as in Eq.~\eqref{eq:alpha}, and the static stiffness is $K_H=I+\hat B^2$. For $C=(I+\beta_*\hat S^2)/\sqrt{1+\beta_*^2}$, the matching condition is $I+\hat B^2=r_*C^\mathsf{T}C$. The transformation $Q=CX$ and $p_Q=C^{-\mathsf{T}}P/2$ defines canonical contact coordinates, with $[Q_n,p_{Q,m}]=i\delta_{nm}$, in which
\begin{equation}
\begin{split}
\frac{H_\mathrm{ad}}{\omega}={}&\Bigl(p_Q+\frac{\hat GQ}{2}\Bigr)^{\!\mathsf{T}}M^{-1}\Bigl(p_Q+\frac{\hat GQ}{2}\Bigr)\\
&+\eta\sum_nw\bigl(Q_n/\sqrt{\eta}\bigr)-\frac N2,
\end{split}\label{eq:Hadcoord}
\end{equation}
where $M=C^{-\mathsf{T}}C^{-1}$ and $\hat G=M\hat B$. The potential is separable, while the kinetic energy couples the contact fields through $M$ and $\hat G$. Since $C$ contains only even powers of $\hat S$, $M$ couples sites within the same sublattice. The matrices $\hat B$ and $C$ commute, giving $\hat G=C^{-\mathsf{T}}\hat BC^{-1}$; hence $\hat G^\mathsf{T}=-\hat G$, and $\hat G$ connects opposite sublattices. The momentum--coordinate term is a gyroscopic coupling with effective vector potential $\mathbf A(Q)=-\hat GQ/2$. Its field strength is
\begin{equation}
\mathcal F_{ij}=\frac{\partial A_j}{\partial Q_i}-\frac{\partial A_i}{\partial Q_j}=\hat G_{ij}.
\end{equation}
For nonzero hopping, $\hat G\neq0$, so this vector potential cannot be removed by a scalar gauge transformation. For either sublattice indicator $v_\mathrm{sub}$, equal to one on that sublattice and zero on the other, $\hat S^2v_\mathrm{sub}=v_\mathrm{sub}$ implies $\hat Bv_\mathrm{sub}=\hat Gv_\mathrm{sub}=0$ and $Mv_\mathrm{sub}=r_*v_\mathrm{sub}$. These relations enter the whole-sublattice tunneling path in Appendix~\ref{app:fullspin}.

\textit{Classical weights.}
In the classical phase-space integral for Eq.~\eqref{eq:Hadcoord}, the substitution $\pi=p_Q+\hat GQ/2$ has unit Jacobian and removes the vector potential from the integrand, as in the Bohr--van Leeuwen theorem. The remaining Gaussian momentum integral is independent of $Q$, giving the factorized contact distribution in Eq.~\eqref{eq:Pcl} with the full local nonlinearity retained. Thus the contact fields are statistically independent and all sign configurations have equal classical weight, while the photon coordinates $X=C^{-1}Q$ remain correlated. As quantum operators, the shifted momenta satisfy $[\pi_i,\pi_j]=i\hat G_{ij}$ and therefore do not form a set of commuting canonical momenta. The vector potential and nondiagonal mass matrix $M$ enter the quantum dynamics and produce intersite harmonic correlations of the contact fluctuations. The cubic vertices then generate the configuration-dependent energy through the three-excitation processes in Eq.~\eqref{eq:Jr}.

\section{Anharmonic corrections and induced Ising coupling}\label{app:kernel}

\textit{Anharmonic vertices.} Expanding the adiabatic spin potential $V(Q)=-\sqrt{\Delta^2+4\lambda^2Q^2}/2$ about $\bar Q_n=\sqrt{\eta}\,\tau_ns_0$ gives
\begin{equation}
V(\bar Q_n)+V'(\bar Q_n)\phi_n+\frac{v_2}{2}\phi_n^2+\tau_n\frac{v_3}{6}\phi_n^3+\frac{v_4}{24}\phi_n^4+\cdots,
\end{equation}
with $v_3$ of Eq.~\eqref{eq:V3} and
\begin{equation}
v_2=-\frac{\omega r_*}{2x^2},\qquad v_4=-\frac{3\omega}{2\eta}\,\frac{r_*^2(4x^2-5)}{x^5}.
\end{equation}
Because $V$ is even in $Q$, even derivatives are the same in both wells and odd derivatives change sign with $\tau_n$. The second-order cubic correction and the first-order quartic correction are both of order $\omega/\eta$. The quartic correction, $(v_4/8)G_0(0)^2N$, is the same for every configuration. Evaluated at the mean-field minima, the leading diagonal Born--Huang correction is $N\omega g^2/(4\eta x^4)$ for every configuration, and retardation of the spin response first shifts the low-energy frequencies at order $\omega/\eta^2$. Neither changes the leading energy differences between configurations.

\textit{Second-order energy.} With $u=\omega t_E$ as in Sec.~\ref{sec:selection}, the second-order energy of $V_3$ in the common harmonic ground state is
\begin{equation}
\begin{split}
E^{(2)}&=-\frac{1}{\omega}\int_0^\infty du\,\langle V_3(u)V_3(0)\rangle_0\\
&=-\frac{v_3^2}{36\,\omega}\sum_{n,m}\tau_n\tau_m\int_0^\infty du\,\langle\phi_n^3(u)\phi_m^3(0)\rangle_0,
\end{split}
\end{equation}
since $\langle V_3\rangle_0=0$. Wick's theorem gives
\begin{equation}
\langle\phi_n^3(u)\phi_m^3(0)\rangle_0=6G_{n-m}(u)^3+9G_0(0)^2G_{n-m}(u).\label{eq:wick}
\end{equation}
The first term connects the two vertices by three propagators and corresponds to intermediate states with three Bogoliubov excitations. The second connects the vertices by one propagator, with one equal-time contraction at each vertex, and corresponds to intermediate states with one excitation. The spectral weight multiplying $e^{iqr}$ in Eq.~\eqref{eq:G} is real, so $G_{-r}(u)=G_r(u)^*$ and the double sum is real.

\textit{One-excitation contraction.} Symmetrizing the integrand of Eq.~\eqref{eq:G} under $q\to-q$ and using $e_q^2-j^2\sin^2q=\mu r_*|F_{\beta_*}(q)|^2$ give
\begin{equation}
\begin{split}
\mathrm{Re}\int_0^\infty du\,G_r(u)&=\int_{-\pi}^{\pi}\frac{dq}{2\pi}\,\cos(qr)\,\frac{|F_{\beta_*}(q)|^2}{e_q^2-j^2\sin^2q}\\
&=\frac{\delta_{r0}}{\mu r_*},
\end{split}
\end{equation}
which is Eq.~\eqref{eq:local}. The one-excitation term therefore contributes $-v_3^2G_0(0)^2N/(4\omega\mu r_*)$, independent of the signs.

\textit{Three-excitation contraction.} The remaining term gives Eq.~\eqref{eq:Epat} with Eq.~\eqref{eq:Jr}. In terms of
\begin{equation}
I_r=2\,\mathrm{Re}\int_0^\infty du\,G_r(u)^3,
\end{equation}
the dimensionless couplings are
\begin{equation}
\mathcal{J}_r(j,x)=\frac{3r_*^3(x^2-1)}{8x^7}\,I_r.
\end{equation}
Since $|F_{\beta_*}(q)|^2$ and $e_q$ are even and $\pi$ periodic, changing variables from $q$ to $\pi-q$ in Eq.~\eqref{eq:G} gives $G_r^*=(-1)^rG_r$. Hence $G_r^3$ is imaginary for odd $r$, and $I_{2m+1}=0$.

\textit{Finite rings.} For the periodic rings considered here, $N$ is a multiple of four. The momentum integral in Eq.~\eqref{eq:G} becomes $N^{-1}\sum_q$ over $q=2\pi k/N$, $k=0,\ldots,N-1$. Equivalently, Poisson summation gives
\begin{equation}
G_r^{(N)}(u)=\sum_{k\in\mathbb Z}G_{r+kN}(u).
\end{equation}
The three-excitation contribution therefore gives the configuration energy
\begin{equation}
\begin{split}
E^{(2)}[\tau]={}&\mathrm{const.}-\frac{v_3^2}{6\omega}\sum_{n\neq m}\tau_n\tau_m\\
&\times\mathrm{Re}\int_0^\infty du\,\bigl[G_{n-m}^{(N)}(u)\bigr]^3.
\end{split}
\end{equation}
The cube includes mixed winding terms from the three propagators. On four sites, only separation two contributes, giving
\begin{equation}
\begin{split}
E^{(2)}[\tau]&=\mathrm{const.}-2J_2^{(4)}(\tau_0\tau_2+\tau_1\tau_3),\\
J_2^{(4)}&=\frac{v_3^2}{6\omega}\,\mathrm{Re}\int_0^\infty du\,\bigl[G_2^{(4)}(u)\bigr]^3.
\end{split}
\end{equation}
This definition preserves the form $-J_2^{(4)}\sum_{n=0}^3\tau_n\tau_{n+2}$, in which each opposite-site pair is counted twice. It accounts for the factor $1/6$ here, compared with $1/3$ in Eq.~\eqref{eq:Jr} for the infinite chain. Relative to FSR, a single defect costs $4J_2^{(4)}$ and CSR costs $8J_2^{(4)}$, giving $C_4^{(4)}=2C_d^{(4)}$. At A, the four-momentum sum gives $C_d^{(4)}=0.01206953$ and hence $\eta J_2^{(4)}/\omega=C_d^{(4)}/4=3.01738\times10^{-3}$, which is 4.14 times the infinite-chain coefficient $\mathcal J_2$.

\textit{Domain walls.} For a reversed block of $L$ consecutive sites on one sublattice of the infinite chain, $2\min(L,m)$ bonds of range $2m$ connect the block to its exterior, each costing $2J_{2m}$. The total energy of the two domain walls is
\begin{equation}
\delta E_\mathrm{pair}(L)=4\sum_{m\ge1}\min(L,m)\,J_{2m}.\label{eq:pair}
\end{equation}
For well-separated walls, $L\to\infty$, this equals $4\sum_mmJ_{2m}=4.004J_2$ at A.

\textit{Numerical evaluation.} At point A, $G_r(u)$ is evaluated using 2048 uniformly spaced momenta and 480 Gauss--Legendre time nodes on $[0,100]$. Refining to 4096 momenta and 800 nodes on $[0,200]$ changes the scaled coefficients $\mathcal{J}_r$, $1\le r\le12$, by at most $1.06\times10^{-13}$. On the refined grid, the nonlocal one-excitation integrals satisfy $|\mathrm{Re}\int du\,G_r|<1.20\times10^{-13}$ for $1\le r\le12$; Fig.~\ref{fig:mechanism}(c) uses their exact zero values. The three-excitation calculation gives
\[
\begin{gathered}
I_2=0.0286835015291,\\
\mathcal{J}_2(A)=7.28264173477\times10^{-4}.
\end{gathered}
\]
The point-B infinite-chain coefficient, evaluated on the refined grid, is $\mathcal{J}_2(B)=1.30675730997\times10^{-4}$. Including $J_{10}$ and $J_{12}$ changes the infinite-chain $\delta E_\mathrm{pair}$ by less than $1.3\times10^{-8}$ in relative terms.

\section{Four-site calculations}\label{app:fullspin}

\subsection{Configuration energies and excitation gaps}

For a configuration $\tau$, the stationary photon displacement $\alpha_\tau$ satisfies $h_\mathrm{ph}\alpha_\tau+\lambda C^\mathsf{T}\langle\sigma^x\rangle_\tau=0$, where $H_\mathrm{ph}=a^\dagger h_\mathrm{ph}a$ and $\langle\sigma_n^x\rangle_\tau=-2\lambda\bar Q_{n,\tau}/(\Delta^2+4\lambda^2\bar Q_{n,\tau}^2)^{1/2}$. Writing $a=\alpha_\tau+c$, the photon basis consists of extended coherent states (displaced Fock states) with $\sum_nm_n\le n_c$, of dimension $\binom{n_c+N}{N}$. With all spin states retained, the total dimension is $2^N\binom{n_c+N}{N}$, giving $117\,040$ states for $N=4$ and $n_c=18$. The displacements of different configurations differ by $O(\sqrt{\eta})$. The lowest Ritz values form a plateau as $n_c$ increases, corresponding to states localized near the chosen mean-field configuration; a complete basis recovers the global spectrum. We extrapolate within this plateau. The FSR and AFSR energies agree within $6\times10^{-10}\omega$, consistent with the degeneracy required by Eq.~\eqref{eq:antiunitary}. Localization is measured by $\delta_\mathrm{loc}=\langle\phi_n^2\rangle_0/(\eta s_0^2)$, which at $\eta=256$ is $0.0138$ at A and $0.0317$ at $x=1.1$.

\begin{table}[ht]
\caption{Intercepts of quadratic fits in $\eta^{-1}$ ($\eta=128$--$1024$) at A, and the four-site analytic values. For the Aitken row, each scaled energy difference $C_\tau(\eta)$ is first extrapolated over $n_c=14,16,18$ at fixed $\eta$; the resulting coefficients are then fitted in $\eta^{-1}$.}
\label{tab:cutoff}
\begin{ruledtabular}
\begin{tabular}{lcc}
Cutoff & $C_d(\infty)$ & $C_4(\infty)$\\
\hline
$n_c=14$ & 0.01203455 & 0.02406938\\
$n_c=16$ & 0.01208161 & 0.02416402\\
$n_c=18$ & 0.01209244 & 0.02418575\\
Aitken & 0.01209757 & 0.02419592\\
Analytic & 0.01206953 & 0.02413905\\
\end{tabular}
\end{ruledtabular}
\end{table}

We use $\eta=128$, 256, 512, and 1024 and $n_c=14$, 16, and 18 for the FSR, one-defect, and CSR configurations. The analytic four-site values follow from the discrete momentum sum of Appendix~\ref{app:kernel}. Figure~\ref{fig:checks}(a) uses Aitken extrapolation in $n_c$ at each $\eta$ (Table~\ref{tab:cutoff}); Fig.~\ref{fig:checks}(b) uses $n_c=18$, and at $\eta=256$ the full-spin and adiabatic gaps there agree within $0.012\%$. The fits use $C_\tau=c_0+c_1/\eta+c_2/\eta^2$. Gaussian averages require an even total number of odd-order vertices, so the energy expansion contains integer powers of $\eta^{-1}$. Quadratic fits to the four values of $\eta$ give intercepts $(C_d,C_4/2)=(0.01209757,0.01209796)$, so that $C_4/C_d=2.0001$. Omitting $\eta=128$ gives $(0.01207459,0.01207466)$ for a quadratic interpolation and $(0.01197089,0.01196999)$ for a linear fit. The corresponding deviations from the analytic value are about $0.24\%$, $0.04\%$, and $0.82\%$. These variations quantify the sensitivity of the extrapolated intercepts to the fit range and order.

For the adiabatic calculations, we use $H_\mathrm{ad}=H_\mathrm{ph}-\sum_n(\Delta^2+4\lambda^2Q_n^2)^{1/2}/2$. We evaluate the square-root operator before projecting onto the cutoff space. A two-mode rotation aligns each normalized contact quadrature with one oscillator, and the square root is then evaluated by 100-node Gauss--Hermite quadrature. Increasing to 120 nodes changes the low energies at $\eta=256$ by less than $4\times10^{-14}\omega$.

\subsection{Tunneling between configurations}

\begin{table}[tbp]
\caption{Four-site saddle actions divided by $\eta$ at $j=0.6$.}
\label{tab:action}
\begin{ruledtabular}
\begin{tabular}{lcc}
Trajectory & $x=1.2$ & $x=1.4$\\
\hline
Single flip & 0.1718 & 0.3605\\
Whole-sublattice flip & 0.1447 & 0.3605\\
\end{tabular}
\end{ruledtabular}
\end{table}

Tunneling adds quantum dynamics to the Ising model~\cite{Kumar2013}, with possible interference between paths~\cite{vonDelft1992}. With $Q=\sqrt{\eta}\,s$ and $u=\omega t_E$, the Euclidean action of Eq.~\eqref{eq:Hadcoord} is
\begin{equation}
S_E=\eta\!\int\!du\Bigl\{\frac14s'^\mathsf{T}Ms'+\sum_n\bigl[w(s_n)-w(s_0)\bigr]-\frac i2s^\mathsf{T}\hat Gs'\Bigr\},
\end{equation}
with saddle equation $-Ms''-2i\hat Gs'+2w'(s)=0$. The action is proportional to $\eta$, so the tunneling magnitudes decay exponentially in $\eta$, with exponent $\mathrm{Re}\,S_E$. The gyroscopic term, which couples the two sublattices, generally makes the saddle trajectories complex, as in magnetic tunneling~\cite{Sharpee2002}. Along a single flip of $s_n$, the neighboring fields $s_{n\pm1}$, which lie on the other sublattice, develop opposite imaginary excursions.

A collective reversal of a uniformly signed sublattice avoids the gyroscopic coupling. We write the trajectory as $s(u)=z(u)v_\mathrm{sub}+s_\mathrm{fixed}$, where $z$ runs from $s_0$ to $-s_0$. The vector $s_\mathrm{fixed}$ has zero components on the moving sublattice and independently chosen values $\pm s_0$ on the other. Since $\hat Gv_\mathrm{sub}=0$ and $Mv_\mathrm{sub}=r_*v_\mathrm{sub}$ (Appendix~\ref{app:quadratic}), the saddle equation reduces exactly to $r_*z''=2w'(z)$; the fixed fields satisfy $w'(\pm s_0)=0$. The gyroscopic term vanishes along this real trajectory. Its action,
\begin{equation}
\frac{S_\mathrm{sub}}{\eta}=\frac N2\sqrt{r_*}\int_{-s_0}^{s_0}\sqrt{w(s)-w(s_0)}\,ds,
\end{equation}
is proportional to the number of reversed fields. On the four-site ring at A, this collective trajectory has a lower action than the single-flip saddle, although it reverses twice as many fields. At B, the real parts of the two saddle-point actions per unit $\eta$ agree to four significant figures, $\mathrm{Re}\,S_E/\eta\simeq0.3605$ (Table~\ref{tab:action}); their near equality is specific to these parameters.

On four sites we solve the saddle equation with the full square-root potentials by central differences and a sparse complex Newton iteration. An integral condition fixes the time-translation zero mode. The grids at A and B have interval lengths 160 and 120 with 4097 and 6145 points, respectively. The discrete residuals are $1.05\times10^{-10}$ and $1.12\times10^{-12}$, and fourth-order continuum checks give $1.23\times10^{-6}$ and $1.92\times10^{-6}$. We evaluate the actions using the discrete variational functional and check the whole-sublattice value against the integral above.

\begin{table}[tbp]
\caption{Tunneling matrix elements of the full-spin Hamiltonian, projected onto orthogonalized localized states, at B for single ($t_1$) and whole-sublattice ($t_\mathrm{sub}$) flips at the largest cutoff $n_c$, with the relative changes of $|t_1|$ and $|t_\mathrm{sub}|$ from $n_c-2$.}
\label{tab:mixing}
\begin{ruledtabular}
\begin{tabular}{ccccc}
$\eta$ & $n_c$ & $|t_1|/\omega$ & $|t_\mathrm{sub}|/\omega$ & Change (\%)\\
\hline
32 & 20 & $1.417\times10^{-5}$ & $4.812\times10^{-5}$ & 0.0264/0.220\\
40 & 22 & $9.084\times10^{-7}$ & $3.081\times10^{-6}$ & 0.0061/0.239\\
48 & 24 & $5.655\times10^{-8}$ & $1.915\times10^{-7}$ & 0.0022/0.217\\
64 & 28 & $2.082\times10^{-10}$ & $7.034\times10^{-10}$ & 0.0005/0.412\\
\end{tabular}
\end{ruledtabular}
\end{table}

To evaluate the four-site amplitudes $t_1\equiv t_1^{(4)}$ and $t_\mathrm{sub}\equiv t_\mathrm{sub}^{(4)}$ of Sec.~\ref{sec:checks}, we compute a dressed local state $|\psi_a\rangle$ about each stationary center in a photon space of cutoff $n_c$, retaining all spin states. From $O_{ab}=\langle\psi_a|\psi_b\rangle$ and $H_{ab}=\langle\psi_a|H|\psi_b\rangle$ we form $H_\mathrm{eff}=O^{-1/2}(H-E_\mathrm{ref}O)O^{-1/2}+E_\mathrm{ref}I$. The off-diagonal element defines the tunneling amplitude $t$, while the difference between the diagonal elements gives the configuration bias. Displacement overlaps use $\langle m|\mathcal{D}(\gamma)|n\rangle=e^{-|\gamma|^2/2}\sqrt{n!/m!}\,\gamma^{m-n}L_n^{(m-n)}(|\gamma|^2)$ for $m\ge n$, with the phase of $\mathcal{D}^\dagger(\alpha_a)\mathcal{D}(\alpha_b)$ retained. In $\langle\psi_a|H|\psi_b\rangle$ we keep the components of $H|\psi_b\rangle$ outside the cutoff space, up to $n_c+1$ total displaced photons, before taking the overlap with $\langle\psi_a|$. The reference energy is subtracted before contraction between centers. The whole-sublattice partner state is generated by the exact antiunitary symmetry of Eq.~\eqref{eq:antiunitary}.

The local eigensolver uses a tolerance of $2\times10^{-14}$ and a Krylov dimension of 48, and the local eigenvector residuals are below $2.4\times10^{-12}\omega$. The cutoff scans start at $n_c=16$ and increase in steps of two up to the largest cutoff in Table~\ref{tab:mixing}. The final cutoff changes are below $0.027\%$ for the single-flip magnitude and $0.42\%$ for the whole-sublattice magnitude. Unweighted fits of $\ln[(|t|/\omega)/\sqrt{\eta}]$ to $\ln a-c\eta$ give $(a,c)=(0.2431,0.3586)$ for the single flip and $(0.8299,0.3588)$ for the whole-sublattice flip. The two decay constants are nearly equal and lie within $0.6\%$ of the corresponding saddle-point actions per unit $\eta$ at B. The larger whole-sublattice amplitude is therefore primarily due to its larger prefactor, with $a_\mathrm{sub}/a_1\simeq3.41$.

\section{Thermal Ising chains and photon observables}\label{app:thermal}

\textit{Transfer matrix.} When tunneling and thermal excitations within the wells are negligible (Sec.~\ref{sec:consequences}), we write $\varsigma_i=\tau_{2i+p}$ for sublattice $p=0,1$. For a uniform contact-ratio mismatch, we use the couplings $\widetilde J_{2m}$ of Eq.~\eqref{eq:Jtilde}, which reduce to $J_{2m}$ at matching. Retaining $m=1,\ldots,R$, the transfer matrix between overlapping blocks is
\begin{equation}
\mathbb{T}_{a,b}=\Bigl[\prod_{\ell=1}^{R-1}\delta_{b_\ell,a_{\ell+1}}\Bigr]\exp\Bigl[\frac{b_R}{k_{\mathrm{B}}T}\sum_{m=1}^R\widetilde J_{2m}a_{R+1-m}\Bigr],
\end{equation}
where $a$ and $b$ are blocks of $R$ consecutive signs and each step advances two physical sites. At matching, we use $R=4$, retaining $J_2$ through $J_8$; $\mathbb{T}$ is then a $16\times16$ matrix, and $Z=[\mathrm{Tr}\,\mathbb{T}^{N/2}]^2$ on an even ring. All allowed transfer energies are shifted by a common constant before exponentiation, and the shift is restored when evaluating the entropy. Thermal corrections to the couplings themselves are exponentially small below the intrawell excitation gap.

\textit{Correlations and entropy.} Let $\Lambda_0$ be the Perron eigenvalue, with left and right eigenvectors normalized by $l_0^\mathsf{T}r_0=1$, and let $\mathcal{S}_{aa}=a_R$. In the thermodynamic limit,
\begin{equation}
\langle\varsigma_0\varsigma_\ell\rangle=\frac{l_0^\mathsf{T}\mathcal{S}\,\mathbb{T}^\ell\mathcal{S}r_0}{\Lambda_0^\ell},\qquad \xi=\frac{2}{\ln(\Lambda_0/|\Lambda_s|)},
\end{equation}
where $\Lambda_s$ is the eigenvalue of largest modulus with nonzero overlap in the sign-odd sector, and the factor of two converts sublattice steps into physical lattice spacings. Since $N^{-1}\ln Z\to\ln\Lambda_0$, the configurational entropy per physical site is $s_\mathrm{conf}/k_{\mathrm{B}}=\ln\Lambda_0+T\,\partial\ln\Lambda_0/\partial T$ at fixed couplings; it excludes the entropy of excitations within the wells. At point A and $k_{\mathrm{B}}T=10J_2$, $s_\mathrm{conf}=0.688k_{\mathrm{B}}$, close to the uncorrelated value $k_{\mathrm{B}}\ln2=0.693k_{\mathrm{B}}$.

\textit{Structure factor.} Define $\mathbb{T}_\perp=\mathbb{T}/\Lambda_0-r_0l_0^\mathsf{T}$, $r_s=\mathcal{S}r_0$, $l_s=\mathcal{S}l_0$, and the generating function
\begin{equation}
\begin{split}
\mathcal{C}(z)&=\sum_{\ell\ge1}z^\ell\langle\varsigma_0\varsigma_\ell\rangle\\
&=l_s^\mathsf{T}\bigl(I-z\mathbb{T}_\perp\bigr)^{-1}z\mathbb{T}_\perp r_s,
\end{split}
\end{equation}
which uses $l_0^\mathsf{T}\mathcal{S}r_0=\langle\varsigma\rangle=0$. Because the two sublattices are independent and equivalent,
\begin{equation}
S_\tau(q)=1+2\,\mathrm{Re}\,\mathcal{C}\bigl(e^{-2iq}\bigr),
\end{equation}
which is $\pi$ periodic and even in $q$. The resolvent avoids truncating the correlation sum. For Fig.~\ref{fig:thermal}(c), we evaluate $W(q)S_\tau(q)$ on 4097 uniformly spaced momenta in the full Brillouin zone, $-\pi\le q\le\pi$, at $k_{\mathrm{B}}T/J_2=1$, 3, and 10.
Since $W(\pi-q)=W(q)$ and $S_\tau(\pi-q)=S_\tau(q)$, the distribution is symmetric about $q=\pi/2$. The side peaks disappear at $k_{\mathrm{B}}T\simeq3.71J_2$, leaving a single maximum at $q=\pi/2$.

\textit{Local photon number.} The local photon number is weighted by $\overline W(q)$ of Appendix~\ref{app:static}. Using the Fourier expansion
\begin{equation}
\frac{1}{1-j^2\sin^2q}=\frac{1}{\sqrt{1-j^2}}\sum_{\ell\in\mathbb Z}(-\beta_*)^{|\ell|}e^{2i\ell q}
\end{equation}
gives
\begin{equation}
\frac{n_\mathrm{loc}^\mathrm{disp}}{\eta}=\frac{s_0^2r_*}{4}\Bigl[\frac{2}{\sqrt{1-j^2}}\bigl(1+2\,\mathcal{C}(-\beta_*)\bigr)-1\Bigr],
\end{equation}
which is used for Fig.~\ref{fig:thermal}(b). The generating function sums the weighted correlations directly, including the contribution of the narrow low-temperature peaks of $S_\tau(q)$. Uncorrelated signs, $\mathcal{C}=0$, give $s_0^2r_*[2/\sqrt{1-j^2}-1]/4=0.1375$ at A. In the zero-temperature limit, the same-sublattice correlations tend to unity, giving $\mathcal{C}(-\beta_*)=-\beta_*/(1+\beta_*)$. The photon number then reaches the FSR/AFSR lower bound $s_0^2r_*/4=0.0917$ of Appendix~\ref{app:static}.

\textit{Numerical values.} At A, the transfer matrix gives $\xi=7.36$ and $0.87$ physical lattice spacings and $n_\mathrm{loc}^\mathrm{disp}/\eta=0.1017$ and $0.1325$ at $k_{\mathrm{B}}T=J_2$ and $10J_2$, respectively. For Figs.~\ref{fig:thermal}(a) and \ref{fig:thermal}(b), we use 241 logarithmically spaced temperatures over $0.2\le k_{\mathrm{B}}T/J_2\le100$.

\section{Deviations from contact matching}\label{app:mismatch}

\textit{Uniform mismatch.} For $\delta\beta=\beta-\beta_*$, the quadratic mean-field term of Eq.~\eqref{eq:EMF} changes by $(\Delta\,\delta\beta/4)\sum_q\partial_\beta\rho_\beta(q)|s_q|^2$. At $\beta_*$,
\begin{equation}
\begin{gathered}
\partial_\beta\rho_\beta(q)\big|_{\beta_*}=d_0+2\sum_{m\ge1}d_m\cos2mq,\\
d_m=-r_*(-\beta_*)^{m-1},
\end{gathered}
\end{equation}
and $\sum_q2\cos(2mq)|s_q|^2=2\sum_ns_ns_{n+2m}$. Inserting $s_n=s_0\tau_n$ gives Eq.~\eqref{eq:Jtilde}, apart from a common onsite shift. The mismatch therefore generates classical couplings with the same even-separation structure and the same geometric tail as the static photon response. At A, the nearest same-sublattice coupling is $\widetilde J_2/\omega=\mathcal{J}_2/\eta+0.1833\,\eta\,\delta\beta$, and the two contributions are equal in magnitude at $|\delta\beta|=6.06\times10^{-8}$ for $\eta=256$, which gives Eq.~\eqref{eq:tolQ}.

The competition between FSR/AFSR and CSR follows from
\begin{equation}
a_j\equiv\partial_\beta\bigl[\rho_\beta(0)-\rho_\beta(\pi/2)\bigr]_{\beta_*}=-\frac{(1+\sqrt{1-j^2})^2(2-j^2)}{2\sqrt{1-j^2}},
\end{equation}
which equals $-3.321$ at $j=0.6$. The energies per site of the FSR and CSR configurations differ by
\begin{equation}
e_U-e_4=\frac{\Delta s_0^2a_j}{4}\,\delta\beta-2\sum_{m\,\mathrm{odd}}J_{2m},
\end{equation}
with $s_0^2a_j/4=-0.3713$ at A. Since $a_j<0$, the mismatch-induced contribution favors FSR/AFSR for positive $\delta\beta$ and CSR for negative $\delta\beta$. The quantum term shifts the boundary between them from $\delta\beta=0$ to $\eta^2\delta\beta=8\mathcal{J}_2/(s_0^2a_j)=-0.00392$ at A, neglecting $J_6$ and longer-range couplings.

\begin{figure}[tbp]
\includegraphics[width=\columnwidth]{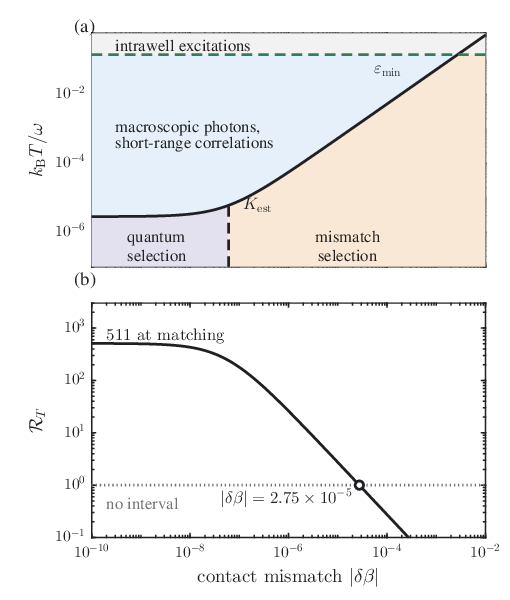}
\caption{Energy scales away from contact matching at A ($j=0.6$, $x=1.2$, $\theta=\pi/2$) and $\eta=256$, versus $|\delta\beta|=|\beta-\beta_*|$, neglecting tunneling. (a) The solid curve is the coupling estimate $K_\mathrm{est}$ and the horizontal dashed line is $\varepsilon_\mathrm{min}$. The vertical dashed line marks equal magnitudes of the quantum and mismatch-induced contributions to $\widetilde J_2$ [Eq.~\eqref{eq:tolQ}]. Shading distinguishes the sign-selection, weakly correlated, and intrawell-excitation regimes; in the mismatch-dominated region, negative $\delta\beta$ favors CSR and positive $\delta\beta$ favors FSR/AFSR. (b) Ratio $\mathcal R_T$ of the upper and lower bounds of the temperature interval, with the factor-of-ten margins of Sec.~\ref{sec:consequences}. The interval closes where $\mathcal R_T=1$, at $|\delta\beta|=2.75\times10^{-5}$ (open circle).}
\label{fig:mismatch}
\end{figure}

\textit{Temperature interval.} For negligible tunneling, the triangle inequality and the geometric series in Eq.~\eqref{eq:Jtilde} give the coupling estimate to first order in $\delta\beta$,
\begin{equation}
\begin{split}
\sum_{m\ge1}|\widetilde J_{2m}|&\lesssim K_\mathrm{est},\\
K_\mathrm{est}&=\sum_{m\ge1}|J_{2m}|+\frac{\Delta s_0^2r_*}{2(1-\beta_*)}|\delta\beta|.
\end{split}
\end{equation}
For the numerical estimates at A, the quantum sum is evaluated using $J_2$--$J_8$, whose omitted tail is negligible on the quoted scale. With the margins $10K_\mathrm{est}<k_{\mathrm{B}}T<0.1\varepsilon_\mathrm{min}$ of Sec.~\ref{sec:consequences}, the interval exists when
\begin{equation}
\mathcal R_T\equiv\frac{0.1\,\varepsilon_\mathrm{min}}{10K_\mathrm{est}}>1.
\end{equation}
At A and $\eta=256$, this ratio is 511 at matching, 179 at $|\delta\beta|=10^{-7}$, 26 at $10^{-6}$, and 2.7 at $10^{-5}$, and it reaches unity at $|\delta\beta|=2.75\times10^{-5}$, corresponding to a relative mismatch $|\delta\beta|/\beta_*\simeq2.5\times10^{-4}$ [Fig.~\ref{fig:mismatch}(b)]. Since the mismatch term in $K_\mathrm{est}$ grows as $\eta\,\delta\beta$ and $\varepsilon_\mathrm{min}$ does not depend on $\eta$, the closing point scales asymptotically as $\eta^{-1}$, whereas Eq.~\eqref{eq:tolQ} scales as $\eta^{-2}$. Figure~\ref{fig:mismatch}(a) compares the relevant energy scales.

\textit{Hopping-phase offset.} For $\theta=\pi/2+\delta\theta$ at fixed $\beta_*$,
\begin{equation}
\delta\rho_\theta(q)=r_*j\cos q\,\frac{1+j^2\sin^2q}{1-j^2\sin^2q}\,\delta\theta+O(\delta\theta^2).
\end{equation}
This correction changes sign under $q\mapsto q+\pi$, so it contains only odd harmonics and generates couplings between signs at odd separations. It breaks the antiunitary symmetry of Eq.~\eqref{eq:antiunitary} and lifts the FSR--AFSR degeneracy:
\begin{equation}
\frac{e_U-e_\mathrm{AFSR}}{\Delta}=\frac{s_0^2r_*j}{2}\,\delta\theta,
\end{equation}
which equals $0.11\,\delta\theta$ at A. This splitting equals $2J_2$ in magnitude at $|\delta\theta|=1.324\times10^{-2}\eta^{-2}$.

\textit{Spatially varying contacts.} For real contact ratios $\beta_n=\beta_*+\delta\beta_n$, let $C_0=(I+\beta_*\hat S^2)/\sqrt{1+\beta_*^2}$ and $C_{nm}=(\delta_{nm}+\beta_n\delta_{m,n+2})/\sqrt{1+\beta_n^2}$, with $K_H=r_*C_0^\mathsf{T}C_0$ fixed. 
\begin{widetext}
\noindent Writing $L=(C-C_0)C_0^{-1}$, we obtain
\begin{equation}
\begin{aligned}
C^{-\mathsf{T}}K_HC^{-1}&=r_*\bigl(I-L^\mathsf{T}-L\bigr)+O(\delta\beta^2),\\
L&=\mathrm{diag}(\delta\beta_n)\Bigl[\sum_{m\ge1}(-\beta_*)^{m-1}\hat S^{2m}-\frac{\beta_*I}{1+\beta_*^2}\Bigr]+O(\delta\beta^2).
\end{aligned}
\end{equation}
\end{widetext}
\begin{samepage}
The onsite contribution is independent of the signs for $s_n=s_0\tau_n$. The remaining terms give the bond correction
\begin{equation}
\begin{split}
\delta H_\mathrm{Ising}={}&-\frac{\Delta r_*s_0^2}{2}\sum_n\sum_{m\ge1}\delta\beta_n\\
&\times(-\beta_*)^{m-1}\tau_n\tau_{n+2m}+\mathrm{const.}
\end{split}\label{eq:disorder}
\end{equation}
\end{samepage}
For $\delta\beta_n=\delta\beta$ this reduces to Eq.~\eqref{eq:Jtilde}. Each atom still couples through real amplitudes to two cavities on the same sublattice, so global parity and the antiunitary symmetry hold in every realization. The leading disorder correction consists of random same-sublattice bonds, while these symmetries forbid random fields. The dominant bonds connect nearest neighbors on each sublattice. The signs of these nearest-neighbor bonds can be removed on an open chain by a local change of the Ising variables $\tau_n$; a ring retains the product of the bond signs as a boundary twist, and the longer-range bonds can introduce frustration.

\bibliography{references}

@article{Schiro2012,
  author  = {Schir{\'o}, M. and Bordyuh, M. and {\"O}ztop, B. and T{\"u}reci, H. E.},
  title   = {Phase transition of light in cavity {QED} lattices},
  journal = {Phys. Rev. Lett.},
  volume  = {109},
  pages   = {053601},
  year    = {2012},
  note    = {Erratum: Phys. Rev. Lett. \textbf{109}, 229901(E) (2012)},
  doi = {10.1103/PhysRevLett.109.053601}
}

@article{Flottat2016,
  author  = {Flottat, T. and H{\'e}bert, F. and Rousseau, V. G. and Batrouni, G. G.},
  title   = {Quantum {Monte Carlo} study of the {Rabi--Hubbard} model},
  journal = {Eur. Phys. J. D},
  volume  = {70},
  pages   = {213},
  year    = {2016},
  doi = {10.1140/epjd/e2016-70492-x}
}

@article{Hwang2015,
  author  = {Hwang, M.-J. and Puebla, R. and Plenio, M. B.},
  title   = {Quantum phase transition and universal dynamics in the {Rabi} model},
  journal = {Phys. Rev. Lett.},
  volume  = {115},
  pages   = {180404},
  year    = {2015},
  doi = {10.1103/PhysRevLett.115.180404}
}

@article{Zhang2021,
  author  = {Zhang, Y.-Y. and Hu, Z.-X. and Fu, L. and Luo, H.-G. and Pu, H. and Zhang, X.-F.},
  title   = {Quantum phases in a quantum {Rabi} triangle},
  journal = {Phys. Rev. Lett.},
  volume  = {127},
  pages   = {063602},
  year    = {2021},
  doi = {10.1103/PhysRevLett.127.063602}
}

@article{Zhao2022,
  author  = {Zhao, J. and Hwang, M.-J.},
  title   = {Frustrated superradiant phase transition},
  journal = {Phys. Rev. Lett.},
  volume  = {128},
  pages   = {163601},
  year    = {2022},
  doi = {10.1103/PhysRevLett.128.163601}
}

@article{FallasPadilla2022,
  author  = {Fallas Padilla, D. and Pu, H. and Cheng, G.-J. and Zhang, Y.-Y.},
  title   = {Understanding the quantum {Rabi} ring using analogies to quantum magnetism},
  journal = {Phys. Rev. Lett.},
  volume  = {129},
  pages   = {183602},
  year    = {2022},
  doi = {10.1103/PhysRevLett.129.183602}
}

@article{Zhao2023,
  author  = {Zhao, J. and Hwang, M.-J.},
  title   = {Anomalous criticality with bounded fluctuations and long-range frustration induced by broken time-reversal symmetry},
  journal = {Phys. Rev. Research},
  volume  = {5},
  pages   = {L042016},
  year    = {2023},
  doi = {10.1103/PhysRevResearch.5.L042016}
}

@article{Li2023,
  author  = {Li, L.-J. and Feng, L.-L. and Dai, J.-H. and Zhang, Y.-Y.},
  title   = {Quantum {Rabi} hexagonal ring in an artificial magnetic field},
  journal = {Phys. Rev. A},
  volume  = {108},
  pages   = {043705},
  year    = {2023},
  doi = {10.1103/PhysRevA.108.043705}
}

@article{Mei2022,
  author  = {Mei, Q.-X. and Li, B.-W. and Wu, Y.-K. and Cai, M.-L. and Wang, Y. and Yao, L. and Zhou, Z.-C. and Duan, L.-M.},
  title   = {Experimental realization of the {Rabi--Hubbard} model with trapped ions},
  journal = {Phys. Rev. Lett.},
  volume  = {128},
  pages   = {160504},
  year    = {2022},
  doi = {10.1103/PhysRevLett.128.160504}
}

@article{Zhu2020,
  author  = {Zhu, G.-L. and Ramezani, H. and Emary, C. and Gao, J.-H. and Wu, Y. and L{\"u}, X.-Y.},
  title   = {Interplay of quantum phase transition and flat band in hybrid lattices},
  journal = {Phys. Rev. Research},
  volume  = {2},
  pages   = {033463},
  year    = {2020},
  doi = {10.1103/PhysRevResearch.2.033463}
}

@article{He2021,
  author  = {He, Y. and Mao, R. and Cai, H. and Zhang, J.-X. and Li, Y. and Yuan, L. and Zhu, S.-Y. and Wang, D.-W.},
  title   = {Flat-band localization in {Creutz} superradiance lattices},
  journal = {Phys. Rev. Lett.},
  volume  = {126},
  pages   = {103601},
  year    = {2021},
  doi = {10.1103/PhysRevLett.126.103601}
}

@article{Gopalakrishnan2009,
  author  = {Gopalakrishnan, S. and Lev, B. L. and Goldbart, P. M.},
  title   = {Emergent crystallinity and frustration with {Bose--Einstein} condensates in multimode cavities},
  journal = {Nat. Phys.},
  volume  = {5},
  pages   = {845--850},
  year    = {2009},
  doi = {10.1038/nphys1403}
}

@article{Strack2011,
  author  = {Strack, P. and Sachdev, S.},
  title   = {{Dicke} quantum spin glass of atoms and photons},
  journal = {Phys. Rev. Lett.},
  volume  = {107},
  pages   = {277202},
  year    = {2011},
  doi = {10.1103/PhysRevLett.107.277202}
}

@article{Buchhold2013,
  author  = {Buchhold, M. and Strack, P. and Sachdev, S. and Diehl, S.},
  title   = {{Dicke}-model quantum spin and photon glass in optical cavities: Nonequilibrium theory and experimental signatures},
  journal = {Phys. Rev. A},
  volume  = {87},
  pages   = {063622},
  year    = {2013},
  doi = {10.1103/PhysRevA.87.063622}
}

@article{Vaidya2018,
  author  = {Vaidya, V. D. and Guo, Y. and Kroeze, R. M. and Ballantine, K. E. and Koll{\'a}r, A. J. and Keeling, J. and Lev, B. L.},
  title   = {Tunable-range, photon-mediated atomic interactions in multimode cavity {QED}},
  journal = {Phys. Rev. X},
  volume  = {8},
  pages   = {011002},
  year    = {2018},
  doi = {10.1103/PhysRevX.8.011002}
}

@article{Guo2019,
  author  = {Guo, Y. and Kroeze, R. M. and Vaidya, V. D. and Keeling, J. and Lev, B. L.},
  title   = {Sign-changing photon-mediated atom interactions in multimode cavity quantum electrodynamics},
  journal = {Phys. Rev. Lett.},
  volume  = {122},
  pages   = {193601},
  year    = {2019},
  doi = {10.1103/PhysRevLett.122.193601}
}

@article{Schuler2020,
 author={Schuler, Michael and De Bernardis, Daniele and L{\"a}uchli, Andreas M. and Rabl, Peter},
 title={The vacua of dipolar cavity quantum electrodynamics},
 journal={SciPost Phys.},volume={9},pages={066},year={2020},doi={10.21468/SciPostPhys.9.5.066}
}

@article{Liang2026,
  author  = {Liang, Y. and Dong, B.-Y. and Xiong, Z. and Zhang, X.-F.},
  title   = {Frustrated {Rydberg} atom arrays meet cavity {QED}: Emergence of the superradiant clock phase},
  journal = {Phys. Rev. Lett.},
  volume  = {136},
  pages   = {073602},
  year    = {2026},
  doi     = {10.1103/m14q-xbzc}
}

@article{Harris1992,
  author  = {Harris, A. B. and Kallin, C. and Berlinsky, A. J.},
  title   = {Possible {N{\'e}el} orderings of the {Kagom{\'e}} antiferromagnet},
  journal = {Phys. Rev. B},
  volume  = {45},
  pages   = {2899--2919},
  year    = {1992},
  doi     = {10.1103/PhysRevB.45.2899}
}

@article{Chubukov1992,
  author  = {Chubukov, A.},
  title   = {Order from disorder in a {kagom{\'e}} antiferromagnet},
  journal = {Phys. Rev. Lett.},
  volume  = {69},
  pages   = {832--835},
  year    = {1992},
  doi     = {10.1103/PhysRevLett.69.832}
}

@article{Barnett2012,
  author  = {Barnett, R. and Powell, S. and Gra{\ss}, T. and Lewenstein, M. and Das Sarma, S.},
  title   = {Order by disorder in spin-orbit-coupled {Bose--Einstein} condensates},
  journal = {Phys. Rev. A},
  volume  = {85},
  pages   = {023615},
  year    = {2012},
  doi = {10.1103/PhysRevA.85.023615},
  note = {Erratum: \href{https://doi.org/10.1103/PhysRevA.85.049905}{Phys. Rev. A \textbf{85}, 049905(E) (2012)}}
}

@article{Yue2021,
  author  = {Yue, M. and Wang, Z. and Mukherjee, B. and Cai, Z.},
  title   = {Order by disorder in frustration-free systems: Quantum {Monte Carlo} study of a two-dimensional {PXP} model},
  journal = {Phys. Rev. B},
  volume  = {103},
  pages   = {L201113},
  year    = {2021},
  doi = {10.1103/PhysRevB.103.L201113}
}

@article{Chernyshev2014,
  author  = {Chernyshev, A. L. and Zhitomirsky, M. E.},
  title   = {Quantum selection of order in an {$XXZ$} antiferromagnet on a {Kagome} lattice},
  journal = {Phys. Rev. Lett.},
  volume  = {113},
  pages   = {237202},
  year    = {2014},
  doi     = {10.1103/PhysRevLett.113.237202}
}

@article{Kockum2014,
  author  = {Kockum, A. F. and Delsing, P. and Johansson, G.},
  title   = {Designing frequency-dependent relaxation rates and {Lamb} shifts for a giant artificial atom},
  journal = {Phys. Rev. A},
  volume  = {90},
  pages   = {013837},
  year    = {2014},
  doi = {10.1103/PhysRevA.90.013837}
}

@article{Kockum2018,
  author  = {Kockum, A. F. and Johansson, G. and Nori, F.},
  title   = {Decoherence-free interaction between giant atoms in waveguide quantum electrodynamics},
  journal = {Phys. Rev. Lett.},
  volume  = {120},
  pages   = {140404},
  year    = {2018},
  doi = {10.1103/PhysRevLett.120.140404}
}

@article{Kannan2020,
  author  = {Kannan, B. and Ruckriegel, M. J. and Campbell, D. L. and Kockum, A. F. and Braum{\"u}ller, J. and Kim, D. K. and Kjaergaard, M. and Krantz, P. and Melville, A. and Niedzielski, B. M. and Veps{\"a}l{\"a}inen, A. and Winik, R. and Yoder, J. L. and Nori, F. and Orlando, T. P. and Gustavsson, S. and Oliver, W. D.},
  title   = {Waveguide quantum electrodynamics with superconducting artificial giant atoms},
  journal = {Nature},
  volume  = {583},
  pages   = {775--779},
  year    = {2020},
  doi = {10.1038/s41586-020-2529-9}
}

@article{Wang2021,
  author  = {Wang, X. and Liu, T. and Kockum, A. F. and Li, H.-R. and Nori, F.},
  title   = {Tunable chiral bound states with giant atoms},
  journal = {Phys. Rev. Lett.},
  volume  = {126},
  pages   = {043602},
  year    = {2021},
  doi = {10.1103/PhysRevLett.126.043602}
}

@article{Du2022,
  author  = {Du, L. and Zhang, Y. and Wu, J.-H. and Kockum, A. F. and Li, Y.},
  title   = {Giant atoms in a synthetic frequency dimension},
  journal = {Phys. Rev. Lett.},
  volume  = {128},
  pages   = {223602},
  year    = {2022},
  doi = {10.1103/PhysRevLett.128.223602}
}

@article{TerradasBrianso2022,
  author  = {Terradas-Brians{\'o}, S. and Gonz{\'a}lez-Guti{\'e}rrez, C. A. and Nori, F. and Mart{\'\i}n-Moreno, L. and Zueco, D.},
  title   = {Ultrastrong waveguide {QED} with giant atoms},
  journal = {Phys. Rev. A},
  volume  = {106},
  pages   = {063717},
  year    = {2022},
  doi = {10.1103/PhysRevA.106.063717}
}

@article{Wang2024,
  author  = {Wang, X. and Zhu, H.-B. and Liu, T. and Nori, F.},
  title   = {Realizing quantum optics in structured environments with giant atoms},
  journal = {Phys. Rev. Research},
  volume  = {6},
  pages   = {013279},
  year    = {2024},
  doi = {10.1103/PhysRevResearch.6.013279}
}

@article{Chen2008,
  author  = {Chen, Q.-H. and Zhang, Y.-Y. and Liu, T. and Wang, K.-L.},
  title   = {Numerically exact solution to the finite-size {Dicke} model},
  journal = {Phys. Rev. A},
  volume  = {78},
  pages   = {051801(R)},
  year    = {2008},
  doi     = {10.1103/PhysRevA.78.051801}
}

@article{Khatua2026,
  author  = {Khatua, S. and Howson, G. C. and Gingras, M. J. P. and Rau, J. G.},
  title   = {Finite-size spectral signatures of order by quantum disorder: A perspective from {Anderson}'s tower of states},
  journal = {Phys. Rev. Lett.},
  volume  = {136},
  pages   = {146702},
  year    = {2026},
  doi     = {10.1103/rdyh-2fqj}
}

@article{Roushan2017,
  author  = {Roushan, P. and Neill, C. and Megrant, A. and Chen, Y. and Babbush, R. and Barends, R. and Campbell, B. and Chen, Z. and Chiaro, B. and Dunsworth, A. and Fowler, A. and Jeffrey, E. and Kelly, J. and Lucero, E. and Mutus, J. and O'Malley, P. J. J. and Neeley, M. and Quintana, C. and Sank, D. and Vainsencher, A. and Wenner, J. and White, T. and Kapit, E. and Neven, H. and Martinis, J.},
  title   = {Chiral ground-state currents of interacting photons in a synthetic magnetic field},
  journal = {Nat. Phys.},
  volume  = {13},
  pages   = {146--151},
  year    = {2017},
  doi = {10.1038/nphys3930}
}

@article{Braumuller2017,
  author  = {Braum{\"u}ller, J. and Marthaler, M. and Schneider, A. and Stehli, A. and Rotzinger, H. and Weides, M. and Ustinov, A. V.},
  title   = {Analog quantum simulation of the {Rabi} model in the ultra-strong coupling regime},
  journal = {Nat. Commun.},
  volume  = {8},
  pages   = {779},
  year    = {2017},
  doi = {10.1038/s41467-017-00894-w}
}

@article{Langford2017,
  author  = {Langford, N. K. and Sagastizabal, R. and Kounalakis, M. and Dickel, C. and Bruno, A. and Luthi, F. and Thoen, D. J. and Endo, A. and DiCarlo, L.},
  title   = {Experimentally simulating the dynamics of quantum light and matter at deep-strong coupling},
  journal = {Nat. Commun.},
  volume  = {8},
  pages   = {1715},
  year    = {2017},
  doi = {10.1038/s41467-017-01061-x}
}

@article{Lv2018,
  author  = {Lv, D. and An, S. and Liu, Z. and Zhang, J.-N. and Pedernales, J. S. and Lamata, L. and Solano, E. and Kim, K.},
  title   = {Quantum simulation of the quantum {Rabi} model in a trapped ion},
  journal = {Phys. Rev. X},
  volume  = {8},
  pages   = {021027},
  year    = {2018},
  doi = {10.1103/PhysRevX.8.021027}
}

@article{Nataf2010,
  author  = {Nataf, P. and Ciuti, C.},
  title   = {No-go theorem for superradiant quantum phase transitions in cavity {QED} and counter-example in circuit {QED}},
  journal = {Nat. Commun.},
  volume  = {1},
  pages   = {72},
  year    = {2010},
  doi = {10.1038/ncomms1069}
}

@article{Viehmann2011,
  author  = {Viehmann, O. and von Delft, J. and Marquardt, F.},
  title   = {Superradiant phase transitions and the standard description of circuit {QED}},
  journal = {Phys. Rev. Lett.},
  volume  = {107},
  pages   = {113602},
  year    = {2011},
  doi = {10.1103/PhysRevLett.107.113602}
}

@article{DeBernardis2018,
  author  = {De Bernardis, D. and Pilar, P. and Jaako, T. and De Liberato, S. and Rabl, P.},
  title   = {Breakdown of gauge invariance in ultrastrong-coupling cavity {QED}},
  journal = {Phys. Rev. A},
  volume  = {98},
  pages   = {053819},
  year    = {2018},
  doi = {10.1103/PhysRevA.98.053819}
}

@article{Sayed2001,
  author  = {Sayed, A. H. and Kailath, T.},
  title   = {A survey of spectral factorization methods},
  journal = {Numer. Linear Algebra Appl.},
  volume  = {8},
  pages   = {467--496},
  year    = {2001},
  doi = {10.1002/nla.250}
}

@article{Ephremidze2009,
  author  = {Ephremidze, L. and Janashia, G. and Lagvilava, E.},
  title   = {A simple proof of the matrix-valued {Fej{\'e}r--Riesz} theorem},
  journal = {J. Fourier Anal. Appl.},
  volume  = {15},
  pages   = {124--127},
  year    = {2009},
  doi = {10.1007/s00041-008-9051-z}
}

@article{Dritschel2004,
  author  = {Dritschel, M. A.},
  title   = {On factorization of trigonometric polynomials},
  journal = {Integral Equations Operator Theory},
  volume  = {49},
  pages   = {11--42},
  year    = {2004},
  doi = {10.1007/s00020-002-1198-4}
}

@article{Kumar2013,
  author  = {Kumar, B. and Jalal, S.},
  title   = {Quantum {Ising} dynamics and {Majorana}-like edge modes in the {Rabi} lattice model},
  journal = {Phys. Rev. A},
  volume  = {88},
  pages   = {011802(R)},
  year    = {2013},
  doi     = {10.1103/PhysRevA.88.011802}
}

@article{vonDelft1992,
  author  = {von Delft, J. and Henley, C. L.},
  title   = {Destructive quantum interference in spin tunneling problems},
  journal = {Phys. Rev. Lett.},
  volume  = {69},
  pages   = {3236--3239},
  year    = {1992},
  doi     = {10.1103/PhysRevLett.69.3236}
}

@article{Sharpee2002,
  author  = {Sharpee, T. and Dykman, M. I. and Platzman, P. M.},
  title   = {Tunneling decay in a magnetic field},
  journal = {Phys. Rev. A},
  volume  = {65},
  pages   = {032122},
  year    = {2002},
  doi     = {10.1103/PhysRevA.65.032122}
}

@misc{GiantAtomZenodo,
  author = {He, Shu and Duan, Li-Wei},
  title = {Data and code for: Extensively degenerate superradiant phase in a giant-atom quantum {Rabi} ring},
  howpublished = {Zenodo, Version 2.0.0},
  year = {2026},
  doi = {10.5281/zenodo.24153863},
  url = {https://doi.org/10.5281/zenodo.24153863},
  note = {\url{https://doi.org/10.5281/zenodo.24153863}}
}

@article{Ashhab2013,
  author={Ashhab, S.},
  title={Superradiance transition in a system with a single qubit and a single oscillator},
  journal={Phys. Rev. A}, volume={87}, pages={013826}, year={2013},
  doi={10.1103/PhysRevA.87.013826}
}

@article{Liu2017,
  author={Liu, M. and Chesi, S. and Ying, Z.-J. and Chen, X. and Luo, H.-G. and Lin, H.-Q.},
  title={Universal scaling and critical exponents of the anisotropic quantum {Rabi} model},
  journal={Phys. Rev. Lett.}, volume={119}, pages={220601}, year={2017},
  doi={10.1103/PhysRevLett.119.220601}
}

@article{Cai2021,
  author={Cai, M.-L. and Liu, Z.-D. and Zhao, W.-D. and Wu, Y.-K. and Mei, Q.-X. and Jiang, Y. and He, L. and Zhang, X. and Zhou, Z.-C. and Duan, L.-M.},
  title={Observation of a quantum phase transition in the quantum {Rabi} model with a single trapped ion},
  journal={Nat. Commun.}, volume={12}, pages={1126}, year={2021},
  doi={10.1038/s41467-021-21425-8}
}

@article{Ye2025,
  author={Ye, T. and Wang, Y.-Z. and Chen, X.-Y. and Chen, Q.-H. and Lin, H.-Q.},
  title={Superradiant phase transitions in the quantum {Rabi} model: Overcoming the no-go theorem through anisotropy},
  journal={Phys. Rev. A}, volume={111}, pages={043716}, year={2025},
  doi={10.1103/PhysRevA.111.043716}
}

@article{Chen2012,
  author={Chen, Q.-H. and Wang, C. and He, S. and Liu, T. and Wang, K.-L.},
  title={Exact solvability of the quantum {Rabi} model using {Bogoliubov} operators},
  journal={Phys. Rev. A}, volume={86}, pages={023822}, year={2012},
  doi={10.1103/PhysRevA.86.023822}
}

@article{Qin2024,
  author={Qin, X. and Zhang, Y.-Y.},
  title={Quantum fluctuations and unusual critical exponents in a quantum {Rabi} triangle},
  journal={Phys. Rev. A}, volume={110}, pages={013713}, year={2024},
  doi={10.1103/PhysRevA.110.013713}
}

@article{FisherSelke1980,
  author  = {Fisher, M. E. and Selke, W.},
  title   = {Infinitely many commensurate phases in a simple {Ising} model},
  journal = {Phys. Rev. Lett.},
  volume  = {44},
  pages   = {1502},
  year    = {1980},
  doi     = {10.1103/PhysRevLett.44.1502},
  note = {Erratum: \href{https://doi.org/10.1103/PhysRevLett.45.148.4}{Phys. Rev. Lett. \textbf{45}, 148(E) (1980)}}
}
\end{document}